\documentclass[a4paper,12pt]{article}
\pdfoutput=1
\usepackage{epsfig}
\usepackage{amssymb}
\usepackage{amsfonts}
\usepackage{amsmath}
\usepackage{euscript}
\usepackage{verbatim}
\usepackage{latexsym}
\usepackage{graphicx}
\usepackage{caption}
\usepackage{float}
\usepackage{subcaption}

\usepackage{listings}

\usepackage{booktabs}

\newif\ifdtup

\catcode`\@=11

\@addtoreset{equation}{section}

\def\@normalsize{\@setsize\normalsize{15pt}\xiipt\@xiipt
\abovedisplayskip 14pt plus3pt minus3pt%
\belowdisplayskip \abovedisplayskip
\abovedisplayshortskip \z@ plus3pt%
\belowdisplayshortskip 7pt plus3.5pt minus0pt}

\def\small{\@setsize\small{13.6pt}\xipt\@xipt
\abovedisplayskip 13pt plus3pt minus3pt%
\belowdisplayskip \abovedisplayskip
\abovedisplayshortskip \z@ plus3pt%
\belowdisplayshortskip 7pt plus3.5pt minus0pt
\def\@listi{\parsep 4.5pt plus 2pt minus 1pt
     \itemsep \parsep
     \topsep 9pt plus 3pt minus 3pt}}

\relax

\catcode`@=12

\catcode`\@=11

\def\section{\@startsection{section}{1}{\z@}{3.5ex plus 1ex minus
   .2ex}{2.3ex plus .2ex}{\large\bf}}

\def\SymBoxes#1#2#3#4{\newdimen\un@t \un@t#3%
\raisebox{#1}{\rule{#2\un@t}{#4}\hskip-#2\un@t
\@tempdimb\un@t \advance\@tempdimb by-#4\@tempcntb#2\relax%
\@whilenum{\@tempcntb>0}\do{
\rule{#4}{\un@t}\hskip\@tempdimb \advance\@tempcntb by\m@ne}%
\hskip-#2\un@t \rule[\un@t]{#2\un@t}{#4}%
\rule[\un@t]{#4}{#4}\hskip-#4
\rule{#4}{\un@t}}\hskip-#4}                

\begin{document}

\newcommand{\beq}{\begin{equation}}
\newcommand{\eeq}{\end{equation}}
\newcommand{\bea}{\begin{eqnarray}}
\newcommand{\eea}{\end{eqnarray}}
\newcommand{\beas}{\begin{eqnarray*}}
\newcommand{\eeas}{\end{eqnarray*}}
\newcommand{\defi}{\stackrel{\rm def}{=}}
\newcommand{\non}{\nonumber}
\newcommand{\bquo}{\begin{quote}}
\newcommand{\enqu}{\end{quote}}
\renewcommand{\(}{\begin{equation}}
\renewcommand{\)}{\end{equation}}
\def \eqn#1#2{\begin{equation}#2\label{#1}\end{equation}}

\def\e{\epsilon}
\def\IZ{{\mathbb Z}}
\def\IR{{\mathbb R}}
\def\IC{{\mathbb C}}
\def\IQ{{\mathbb Q}}
\def\de{\partial}
\def\Tr{ \hbox{\rm Tr}}
\def\H{ \hbox{\rm H}}
\def\HE{ \hbox{$\rm H^{even}$}}
\def\HO{ \hbox{$\rm H^{odd}$}}
\def\K{ \hbox{\rm K}}
\def\Im{ \hbox{\rm Im}}
\def\Ker{ \hbox{\rm Ker}}
\def\const{\hbox {\rm const.}}
\def\o{\over}
\def\im{\hbox{\rm Im}}
\def\re{\hbox{\rm Re}}
\def\bra{\langle}\def\ket{\rangle}
\def\Arg{\hbox {\rm Arg}}
\def\Re{\hbox {\rm Re}}
\def\Im{\hbox {\rm Im}}
\def\exo{\hbox {\rm exp}}
\def\diag{\hbox{\rm diag}}
\def\longvert{{\rule[-2mm]{0.1mm}{7mm}}\,}
\def\a{\alpha}
\def\dag{{}^{\dagger}}
\def\tq{{\widetilde q}}
\def\p{{}^{\prime}}
\def\W{W}
\def\N{{\cal N}}
\def\hsp{,\hspace{.7cm}}

\def\br{\nonumber}
\def\IZ{{\mathbb Z}}
\def\IR{{\mathbb R}}
\def\IC{{\mathbb C}}
\def\IQ{{\mathbb Q}}
\def\IP{{\mathbb P}}
\def \eqn#1#2{\begin{equation}#2\label{#1}\end{equation}}

\newcommand{\C}{\ensuremath{\mathbb C}}
\newcommand{\Z}{\ensuremath{\mathbb Z}}
\newcommand{\R}{\ensuremath{\mathbb R}}
\newcommand{\rp}{\ensuremath{\mathbb {RP}}}
\newcommand{\cp}{\ensuremath{\mathbb {CP}}}
\newcommand{\vac}{\ensuremath{|0\rangle}}
\newcommand{\vact}{\ensuremath{|00\rangle}                    }
\newcommand{\oc}{\ensuremath{\overline{c}}}
\newcommand{\psizero}{\psi_{0}}
\newcommand{\phizero}{\phi_{0}}
\newcommand{\hzero}{h_{0}}
\newcommand{\psiin}{\psi_{\rh}}
\newcommand{\phiin}{\phi_{\rh}}
\newcommand{\hin}{h_{\rh}}
\newcommand{\rh}{r_{h}}
\newcommand{\rb}{r_{b}}
\newcommand{\psibnd}{\psi_{0}^{b}}
\newcommand{\psibndp}{\psi_{1}^{b}}
\newcommand{\phibnd}{\phi_{0}^{b}}
\newcommand{\phibndp}{\phi_{1}^{b}}
\newcommand{\gbnd}{g_{0}^{b}}
\newcommand{\hbnd}{h_{0}^{b}}
\newcommand{\zh}{z_{h}}
\newcommand{\zb}{z_{b}}
\newcommand{\man}{\mathcal{M}}
\newcommand{\hbr}{\bar{h}}
\newcommand{\tbr}{\bar{t}}

\begin{titlepage}

\def\thefootnote{\fnsymbol{footnote}}

\begin{center}
{\large
{\bf A Smooth Horizon without a Smooth Horizon
}
}
\end{center}

\begin{center}
Vaibhav Burman$^a$\footnote{\texttt{vaibhav2021@iisc.ac.in  }}, \ Suchetan Das$^a$\footnote{\texttt{suchetan1993@gmail.com }}, Chethan Krishnan$^a$\footnote{\texttt{chethan.krishnan@gmail.com}} \  

\end{center}

\renewcommand{\thefootnote}{\arabic{footnote}}

\begin{center}

$^a$ {Center for High Energy Physics,\\
Indian Institute of Science, Bangalore 560012, India}\\

\end{center}
\vspace{-0.15in}
\noindent
\begin{center} {\bf Abstract} \end{center}
Recent observations on type III algebras in AdS/CFT raise the possibility that smoothness of the black hole horizon is an emergent feature of the large-$N$  limit. In this paper, we present a {\em bulk} model for the finite-$N$ mechanism underlying this transition. We quantize a free scalar field on a BTZ black hole with a Planckian stretched horizon placed as a Dirichlet boundary for the field. This is a tractable model for the stretched horizon that does not ignore the angular directions, and it defines a black hole vacuum which has similarities to (but is distinct from) the Boulware state. Using analytic approximations for the normal modes, we first improve upon 't Hooft's brick wall calculation: we are able to match {\em both} the entropy and the temperature, {\em exactly}. Emboldened by this, we compute the boundary Wightman function of the scalar field in a typical pure state built on our stretched horizon vacuum, at an energy sliver at the mass of the black hole. A key result is that despite the manifest lack of smoothness, this single-sided pure state calculation yields precisely the Hartle-Hawking thermal correlator associated to the smooth horizon, in the small-$G_N$ limit. At finite $G_N$, there are variance corrections that are suppressed as  $\mathcal{O}(e^{-S_{BH}/2})$. They become important at late times and resolve Maldacena's information paradox. Highly excited typical pure states on the stretched horizon vacuum are therefore models for black hole microstates, while the smooth horizon describes the thermal state. We note that heavy excited states on the stretched horizon are better defined than the vacuum itself. These results suggest that complementarity in the bulk EFT could arise from a UV complete bulk description in which the black hole interior is not manifest.

\vspace{1.6 cm}
\vfill

\end{titlepage}

\setcounter{footnote}{0}

\tableofcontents

\section{Introduction}

Classical black holes in bulk effective field theory (EFT) have smooth horizons, and there are extremely good reasons to think that smooth horizons carry entropy. However, in bulk EFT, there is no mechanism to account for this entropy. The bulk description of black hole microstates is therefore an outstanding question for quantum gravity. 

Recent developments \cite{Leutheusser1, Leutheusser2} suggest that in the large-$N$ limit\footnote{In AdS/CFT, the ratio between curvature radius $L$ of AdS and the Planck length $l_p$ is a positive power of the gauge group rank $N$. In the AdS$_3$/CFT$_2$ setting that we will be working with, this takes the famous Brown-Henneaux \cite{Brown-Henneaux} form $c=\frac{3L}{2G_N}$. The central charge $c$ can be viewed as the AdS$_3$ analogue of a (positive power of) $N$.}, the algebra of fluctuations around a heavy black hole state in a holographic CFT becomes type III (instead of being type I, which is the case at finite $N$). Type III algebras have non-trivial commutants, which indicates that new causal structures can emerge in this limit. If true, we come across a remarkable possibility  -- the black hole interior is an emergent feature of the large-$N$ limit, and at finite-$N$ the interior only makes sense in an approximate sense. An analogy that is useful to keep in mind  is that between this large-$N$ phase transition and the large-volume transitions in magnets (or thermodynamics, more generally). Phase transitions require infinite volumes, but magnets still make sense as extremely good approximations as long as the volumes are macroscopic. In other words, making the large-$N$ limit precise as a well-defined scaling limit is likely key for understanding the black hole interior at finite-$N$. 

A complete understanding of Yang-Mills theories at (large, but) finite-$N$ is believed to be an exceptionally difficult problem \cite{Jaffe}\footnote{Matrix models with quantum gravity duals \cite{BFSS} may be the simplest setting to make progress, but even there progress so far has been via numerical lattice methods \cite{Hanada}.}. So it may be worthwhile trying to reach finite-$N$ from the ``other end'', ie., by starting with the bulk gravitational theory (understood perturbatively around $N = \infty$) and trying to guess the modifications that happen when $N$ is finite. Clearly, this is also a highly non-trivial task, precisely because of the existence of the large-$N$ phase transition. 

In this paper, we will present evidence that the stretched horizon \cite{tHooft, STU, Mathur} should be interpreted as a bulk model for a finite-$N$ UV-complete description of a black hole microstate\footnote{More precisely, we will argue that black hole microstates are modeled by excitations on the stretched horizon at an energy sliver at the mass of the black hole.}. We will use the phrases brickwall \cite{tHooft}, stretched horizon \cite{STU} and fuzzball \cite{Mathur} interchangeably in this paper. The unifying theme here for our purposes is that these words all refer to a bulk description of the black hole that does not have a manifest interior. We will adopt a specific interpretation of this picture -- that this absence of the interior is a statement about the UV complete bulk description at finite-$N$. Being specific about this, allows us to have a sharp setting to do calculations in, and find evidence for or against such a picture. We do not claim that the authors of \cite{tHooft, STU, Mathur} do (or do not) share our views on this specific interpretation. Our perspective is closest in spirit to some of the comments in the Introduction of \cite{WittenCrossed}.

Our goal is to use this picture to make some statements about the emergence of bulk EFT at the black hole horizon. Far from the horizon, the distinction between bulk EFT and the UV complete description is unimportant. Therefore, away from the horizon we will simply use the bulk EFT itself to stand in for the UV complete description\footnote{We emphasize that this is clearly an approximation, especially at finite-$N$. Of course we need a cut-off to make sense of any EFT. However, we do not expect this to affect discussions of the black hole horizon/interior. Whatever method one uses to introduce and deal with EFT cut-offs should work fine.}. But we expect the emergence of the bulk EFT description from the UV complete description to be more subtle, starting about a Planck length outside the horizon. We will need a better model for the UV complete description than simply the bulk EFT, if we expect the UV complete description to model microstates. This is because bulk EFT will carry unexplained entropy if it is used to describe the horizon, and microstates (by definition) cannot carry entropy. Motivated by our previous discussions, we will view the brickwall boundary condition as a proxy for such a UV complete description. In other words, the brick wall is a UV regulator for bulk EFT. 

With these preliminary remarks to set up the context, the question of the emergence of a smooth horizon from the UV-complete description becomes: 

{\em What is the EFT description associated to a bulk EFT, but with a (Planckian) stretched horizon?} 

\noindent
We will present strong evidence that the natural effective correlators of this set up are, in a precise sense, the correlators associated to a smooth horizon. Along the way, we find other revealing results, some of which we outline below (loosely in the order in which they appear in the text). See also the discussion in the two concluding sections.
\begin{itemize}
\item We quantize a scalar field with a stretched horizon boundary condition on the BTZ black hole. This defines quantum field theory in the BTZ background built on a stretched horizon vacuum, and defines its associated correlators. The finiteness of the stretched horizon radius is a measure of the finiteness of $G_N$ or $N$, which we make precise.
\item The energy spectrum of these solutions is quantized -- these  are normal modes as opposed to quasi-normal modes. We use the Klein-Gordon inner product to fix the normalizations of the wave equation solutions. This normalization depends on the stretched horizon location, and is a crucial technical ingredient in our calculation.
\item We can argue that it is a certain low-lying part of the spectrum that is responsible for the thermodynamics of the black hole. In this regime, we can determine the normal modes analytically. Using them we reproduce the entropy and temperature of the BTZ black hole exactly (see also \cite{Pradipta}), once we specify the mass of the black hole. In other words, this is a conventional statistical mechanics calculation. 
\item A crucial feature of the normal mode spectrum is that there is an approximate degeneracy in the angular directions in the low-lying part of the spectrum. Furthermore, the calculation naturally suggests a cut-off in the angular quantum numbers, but the cut-off is  a few orders of magnitude smaller than the cut-off suggested by 't Hooft for semi-classical trapping reasons. This is due to the presence of a $\sim \log S_{BH}$ factor. Details of this are discussed in \cite{Pradipta}. There are multiple reasons to think that this refined cut-off is more natural -- the evidence for this goes beyond the exact match of the entropy and temperature mentioned above.   
\item We compute the two point correlation function of the scalar field in a microcanonical typical {\em pure} state\footnote{By this, we mean a pure state chosen randomly from a suitable (Haar) distribution, from the Hilbert space of states in the microcanonical sliver. Note that this need not be an eigenstate.  We will have more to say about this in Section \ref{sup-var}.} built on this stretched horizon vacuum, at an energy sliver equal to the mass of the black hole. We view such microstates as models for black hole microstates. This calculation and the quantum field theory on the stretched horizon background on which it is built, are the key technical developments in this paper. Some results on the determination of normal modes have appeared elsewhere \cite{Pradipta, Adepu, Sumit1, Riemann,  Arnab}. 
\item The result of this calculation shows that in the $N \rightarrow \infty$ limit, the two point Wightman correlator on this pure state\footnote{In fact, the statements in this bullet point are true also for generic eigenstates. It is only to argue {\em exponential} variance suppression in the next bullet point, that we need microcanonical typical states.} reduces precisely to the Hartle-Hawking thermal correlator associated to the smooth horizon at the correct Hawking temperature. All the corrections vanish in this limit. In other words, the smooth horizon correlator is emergent.  
\item At finite-$N$, there are variance corrections\footnote{We emphasize that we use ensembles of pure states only to argue typicality, and there are no explicit ensemble averages of any physical quantities anywhere in this paper. See \cite{Vyshnav-Page} for reasons to think that the ensembles that have been noted in recent discussions of the Page curve, are likely emergent and not fundamental.} that are suppressed  by $e^{-S_{BH}/2}$ which become important at late times of $\sim \mathcal{O}(S_{BH})$. This can be viewed as a stretched horizon resolution of Maldacena's information paradox \cite{Maldacena}. 
\item A crucial observation then, is that the order of $t \rightarrow \infty$ vs $N \rightarrow \infty$ matters. That this should be the case, has been suggested by some of the recent discussions on von Neumann algebras and AdS/CFT  \cite{Lashkari}. The stretched horizon provides a mechanism for this to be realized. 
\item It is worth noting that our calculation ``cancels'' two potentially problematic features against each other. Firstly, it is known that the stress tensor vev near the horizon diverges with the (stretched horizon) radius. This can be viewed as the price to be paid for installing the stretched horizon and violating the principle of equivalence\footnote{Note however that our results are ultimately consistent with the possibility that the stretched horizon is a UV cut-off that is invisible in the emergent bulk EFT.}. Secondly, excitations on the stretched horizon vacuum that are heavy enough to contribute to the black hole mass, are problematic. We do not want fluctuations that are heavy enough to backreact when considering quantum field theory in curved spacetime. Interestingly, there is a sense in which these two ``errors'' annihilate each other when you put them together -- note in particular that the stress tensor vev is negative\footnote{It is remarkable that an object that in many ways is naturally interpreted as a UV regulator, has this striking property. We suspect this is a consequence of the fact that UV-completion in gravity is inextricably tied to regulating $N$ away from infinity.}. The observation of this cancellation was first made by Israel and Mukohyama \cite{Israel-Mukohyama} long ago. They had a stretched horizon, but they worked with the Boulware vacuum to make their quasi-heuristic claims, and not the genuine stretched horizon vacuum. Our stretched horizon vacuum is the correct setting to interpret their prescient observations, and the argument is identical. 
\end{itemize}

\section{Scalar Field in BTZ and Normal Modes}

In this paper, we will consider a scalar field $\Phi(r,t,\psi)$ in the background of the 2+1-dimensional non-rotating BTZ black hole whose metric is written as
\begin{align}\label{BTZ metric1}
ds^2 = -\frac{(r^2-r_{h}^{2})}{L^2}dt^2 + \frac{L^2}{(r^2-r_{h}^{2})}dr^2 + r^2d\psi^2
\end{align}
where $r_{h}$ is the horizon radius and $L$ is the AdS radius. Most of the discussion in this section is review of \cite{Adepu, Sumit1}, but we present some details to set the context and notation. Some of this material is also crucial for the discussions in Appendix \ref{NormApp} which is an important technical ingredient in our work.

The scalar field $\Phi$ satisfies the wave equation
\begin{align}\label{Wave equation1}
    \frac{1}{\sqrt{|g|}}\partial_{\mu}\left(\sqrt{|g|}g^{\mu\nu}\partial_{\nu}\Phi\right) = m^2\Phi
\end{align}
Since the metric \eqref{BTZ metric1} has time translational and rotational symmetry, equation \eqref{Wave equation1} can be solved by writing $\Phi(r,t,\phi) = \frac{1}{\sqrt{r}}\sum_{\omega,J}e^{-i\omega t}e^{iJ\psi}\phi_{\omega,J}(r)$. We get the second-order differential equation for the radial part of the field
\begin{align}\label{Radial DE}
    (r^2-r_{h}^2)\phi_{\omega,J}^{''}(r)+2r(r^2-r_{h}^2)\phi_{\omega,J}^{'}(r)+\omega^{2}L^{4}\phi_{\omega,J}(r)-V_{J}(r)\phi_{\omega,J}(r) = 0
\end{align}

\noindent Here $V_{J}(r) = (r^2-r_{h}^2)\left[\frac{1}{r^{2}}\left( J^{2}L^{2}+\frac{r_{h}^{2}}{4} \right) + \nu^{2}-\frac{1}{4}  \right] $ and $\nu = \sqrt{1+m^2L^2}$.
The solution of the above equation is 
\begin{align}\label{Radial DE solution}
    \phi_{\omega,J}(r) = r^{\frac{1}{2}-\frac{iJL}{r_{h}}}(r^2-r_{h}^2)^{-\frac{i\omega L^{2}}{2r_{h}}}\left[e^{-\frac{\pi JL}{r_{h}}}\big(\frac{r}{r_{h}}\big)^{\frac{2iJL}{r_{h}}} C_{2}H(r) + C_{1}G(r) \right]
\end{align}
where, 
\begin{align}
    G(r) = {}_{2}F_{1}\left[\frac{1}{2}\left(1-\frac{i\omega L^{2}}{r_{h}}-\frac{iJL}{r_{h}}-\nu  \right), \frac{1}{2}\left(1-\frac{i\omega L^{2}}{r_{h}}-\frac{iJL}{r{h}}+\nu  \right);1-\frac{iJL}{r_{h}},\frac{r^{2}}{r_{h}^{2}} \right]
\end{align}
\begin{align}
    H(r) = {}_{2}F_{1}\left[\frac{1}{2}\left(1-\frac{i\omega L^{2}}{r_{h}}+\frac{iJL}{r_{h}}-\nu  \right), \frac{1}{2}\left(1-\frac{i\omega L^{2}}{r_{h}}+\frac{iJL}{r_{h}}+\nu  \right);1+\frac{iJL}{r_{h}},\frac{r^{2}}{r_{h}^{2}} \right]
\end{align}
Expanding the equation (\ref{Radial DE solution}) at $r\rightarrow\infty$(boundary) and demanding the normalizability, we get 
\begin{align}
    C_{2}=-\frac{\gamma(J,\nu)}{\gamma(-J,\nu)}C_{1}
\end{align}
where 
\bea
\gamma(J,\nu)\equiv\frac{\Gamma(\nu)\Gamma(1-\frac{iJL}{r_{h}})}{\Gamma\left[\frac{1}{2}\left(1-\frac{i\omega L^{2}}{r_{h}}-\frac{iJL}{r_{h}}+\nu  \right)\right]\Gamma\left[\frac{1}{2}\left(1+\frac{i\omega L^{2}}{r_{h}}-\frac{iJL}{r_{h}}+\nu  \right)\right]}.
\eea
The above solution can be written in terms of a single hypergeometric function by using hypergeometric identities
and the radial solution (\ref{Radial DE solution}) becomes 
\begin{align}\label{normalized at boundary}
\phi_{\omega,J}(r) = S r^{\frac{1}{2}-\frac{iJL}{r_{h}}}(r^{2}-r_{h}^{2})^{-\frac{i\omega L^{2}}{2r_{h}}}e^{-i\pi\beta} 
\left(\frac{r}{r_{h}}\right)^{-2\beta}{}_{2}F_{1}
\left[\alpha,\beta;\Delta;y\right]
C_{1}
\end{align}
where we have defined 
\bea
& S=\frac{\Gamma\left(\frac{1}{2}(1+\frac{i\omega L^{2}}{r_{h}}+\frac{iJL}{r_{h}}+\nu)\right)\Gamma\left(\frac{1}{2}(1-\frac{i\omega L^{2}}{r_{h}}+\frac{iJL}{r_{h}}+\nu)\right)}{\Gamma(1+\nu)\Gamma\left(\frac{iJL}{r_{h}}\right)}; \ \ \Delta = 1+\nu; \ \ y = \frac{r_{h}^{2}}{r^{2}}; \\ 
& \alpha = \frac{1}{2}\left(1-\frac{i\omega L^{2}}{r_{h}}+\frac{iJL}{r_{h}}+\nu\right); \ \ \beta = \frac{1}{2}\left(1-\frac{i\omega L^{2}}{r_{h}}-\frac{iJL}{r_{h}}+\nu\right).
\eea 

Now we demand that the field $\phi_{\omega,J}(r)$ satisfies a Dirichlet boundary condition at the stretched horizon ($r=r_{h}+\epsilon,\epsilon>0$). This means that 
\begin{align}\label{Dirichlet condition}
    \phi_{\omega,J}(r=r_{h}+\epsilon) = 0
\end{align}
This boundary condition gives us a quantization condition on the modes ($\omega$'s). So instead of writing $\omega$, we will sometimes write $\omega_{n,J}$. To get the quantization condition we expand (\ref{normalized at boundary}) around $r=r_{h}$. It is possible to show that this is a reliable approximation in the regimes we will be interested in \cite{Pradipta}. If we write (\ref{normalized at boundary}) in terms of $\epsilon$ then we have to do the expansion around $\epsilon=0$. Let's write $r = r_{h}+\epsilon $,  then (\ref{normalized at boundary}) becomes
\begin{align}\label{normalized at boundary 2}
&\hspace{-5mm}\phi_{\omega,J}(r_{h}+\epsilon) = C_{1}S \left(1+\frac{\epsilon}{r_{h}}\right)^{-\frac{1}{2}+\frac{i\omega L^{2}}{r_{h}}-\nu}\hspace{-2mm}r_{h}^{\frac{1}{2}-\frac{iJL}{r_{h}}-\frac{i\omega L^{2}}{r_{h}}}\hspace{-2mm}\left(\frac{2\epsilon}{r_{h}}\right)^{-\frac{i\omega L^{2}}{r_{h}}}\hspace{-2mm}e^{-i\pi\beta}{}_{2}F_{1}\left[\alpha,\beta;\Delta;1-\frac{2\epsilon}{r_{h}}\right]
\end{align}
Doing the series expansion around $\epsilon=0$, we get
\begin{align}
    \phi_{hor}(r)\approx C_{1}\left(P_{1}\left(\frac{\epsilon}{r_{h}}\right)^{-\frac{i\omega L^{2}}{2r_{h}}}+Q_{1}\left(\frac{\epsilon}{r_{h}}\right)^{\frac{i\omega L^{2}}{2r_{h}}}\right) \label{near-hor-exp}
\end{align}
where $P_{1}$ and $Q_{1}$ are defined in Appendix \ref{StretchExp}. Since $\phi_{hor}(r_{h}+\epsilon) = 0$, we get
\begin{align}\label{quantization condition}
    \frac{P_{1}}{Q_{1}} = -\left(\frac{\epsilon}{r_{h}}\right)^{\frac{i\omega L^{2}}{r_{h}}}
\end{align}
We will use the above equation to find the quantization condition on $\omega's$.

For a massless scalar field (ie., $\nu=1$ and $m=0$) the ratio $\frac{P_{1}}{Q_{1}}$ reduces to\footnote{In the next two expressions, we are setting $\frac{\omega_{n,J}L^{2}}{r_{h}}\rightarrow \omega$ and $\frac{JL}{r_{h}}\rightarrow J$ to avoid too much clutter.}
\begin{align}\label{P1/Q1}
\frac{P_{1}}{Q_{1}} &\equiv 2^{-i\omega}\frac{e^{\pi J}-e^{-\pi \omega}}{e^{\pi J}-e^{\pi \omega}} 
\frac{\Gamma\left(-\frac{i}{2}(\omega+J)\right)\Gamma\left(1-\frac{i}{2}(\omega+J)\right)\Gamma(i\omega)}{\Gamma\left(\frac{i}{2}(\omega-J)\right)\Gamma\left(1+\frac{i}{2}(\omega-J)\right)\Gamma(-i\omega)} \ e^{\pi \omega}
\end{align}
This quantity is a pure phase, so taking argument on both sides of \eqref{quantization condition}, we obtain
\begin{align}\label{dispersion eqtn}
\hspace{-2mm}2{\rm Arg}(\Gamma(i\omega))+2{\rm Arg}(\Gamma(-\frac{i}{2}(\omega+J)))+2{\rm Arg}(\Gamma(\frac{i}{2}(J-\omega))) = \omega\log\Big(\frac{2\epsilon}{r_{h}}\Big)+(2n-1)\pi
\end{align}
where $n$ is a free integer. We will not present the details of the solution \cite{Pradipta} -- but it turns out that in the low-lying regime\footnote{By low-lying here, we simply mean low $J$ with respect to $J_{max}$ (defined below). The $n$-dependence is in fact exactly linear in the near-horizon approximation \eqref{near-hor-exp} even beyond the low-lying $J$ approximation. In fact, we have checked \cite{Pradipta} that the $n$-dependence is linear even if we retain the next order corrections in $\epsilon$ beyond \eqref{near-hor-exp}. Because of the structure of the equations that emerge, we suspect this could even be an all order result. We have no statements (yet) about potential corrections in $n$-dependence that are non-perturbative in $\epsilon$. Let us also comment that in the Rindler case \cite{Pradipta}, the exact radial normal mode condition $\phi(r_h+\epsilon)=0$ is exactly solvable numerically, and they match the qualitative features of BTZ that we find here.} (which is the one of interest to us) these normal modes can in fact be analytically solved. One expression that we will present is 
\begin{align}
     \omega_{n,J}\sim -\frac{2n\pi r_{h}}{L^{2}\Big(\log\Big(\frac{2\epsilon}{r_{h}}\Big)+\log\Big(\frac{JL}{r_{h}}\Big)^{2}\Big)}
\end{align}
There are some small corrections to this expression even at low $\omega$ which slightly change the $J$-dependence \cite{Pradipta}. But it turns out that they do not affect our results -- nor do they affect the qualitative statement that the low-lying analytic expression for $\omega_{n,J}$ undergoes a complete breakdown for some $J_{max}$. The precise value of $J_{max}$ is not too important, we will take it to be $\frac{r_{h}^{3/2}}{L\sqrt{2\epsilon}}$ which is where the above expression diverges. Because of the smallness of $\epsilon$, the contribution of the $\log J$ term is negligible compared to the $\log\Big(\frac{2\epsilon}{r_{h}}\Big)$ term, when $J$ is not too large. So the final expression of the normal modes that will matter to us, is
\begin{align}\label{normal modes}
    \omega_{n,J}\approx -\frac{2n\pi r_{h}}{L^{2}\log\Big(\frac{2\epsilon}{r_{h}}\Big)}
\end{align}
A version of this expression was written down by Solodukhin \cite{Solodukhin}\footnote{See also the related work in \cite{Debajyoti}.}, see his eqn (3.8). This formula is of importance to us because together with a few other significant inputs, it gives us the black hole entropy, Hawking temperature and the Hartle-Hawking correlator -- all on the nose, as we are going to see in the next sections. A crucial feature about this equation that we will put to good use is that it is degenerate in $J$.

\section{Partition Function, Entropy and Density of states}

In this section we will study statistical mechanics of the low-lying normal modes. We will compute the partition function, density of states and entropy associated to the Fock space constructed out of the quantized free scalar field on the stretched horizon background. 

By counting the modes semi-classically trapped behind the angular momentum barrier,  't Hooft obtained the area scaling of the  entropy \cite{tHooft}. To do this, he used an indirect approach where he did not explicitly compute the normal modes. Because of this, he needed to input both the mass and the temperature of the black hole. Since we have the explicit normal modes, we will need to only specify the mass -- this specifies the ensemble, and fixes all the other thermodynamic quantities. In this sense, our calculation is directly analogous to Planck's calculation except that our normal modes have a degeneracy in the angular direction of the black hole -- in the Planckian set up, there is linear harmonic oscillator growth in the quantum numbers associated to each of the spatial directions. This degeneracy is the underlying mechanism behind the area-scaling of the entropy \cite{Pradipta} as opposed to the volume-scaling in the Planckian setting.

The effect of 't Hooft's semi-classical trapping argument is to provide a cut-off in $J$ \cite{Pradipta}. We will simply fix the cut-off by demanding that it leads to the correct Bekenstein-Hawking entropy. Remarkably, we find that this leads to exactly the correct Hawking temperature as well. Interestingly, this brings down the $J_{cut}$ to a regime where the low-lying analytic spectrum is in fact, reliable. More discussion on this improved $J_{cut}$ and other details can be found in \cite{Pradipta}. A version of the calculation we present here that generalizes to rotating black holes can also be found there \cite{Pradipta}.
 
We start by defining a canonical partition function to compute density of states.  The partition function of the fields in thermal equilibrium with some temperature $\beta_{SH}$ can be expressed as
\begin{align}
Z_{SH}(\beta_{SH})= Tr(e^{-\beta_{SH} :H:}) = \prod_{n,J}\sum_{N_{n,J}}e^{-\beta_{SH}\omega_{n,J}N_{n,J}} = \prod_{n,J}\frac{1}{1-e^{-\beta_{SH}\omega_{n,J}}}
\end{align}
where the normal ordered Hamiltonian $:H: \equiv \sum_{n,J}\omega_{n,J}b^{\dagger}_{n,J}b_{n,J}$. Hence,
\begin{align}
\log Z_{SH} = - \sum_{n,J}\log\left(1-e^{-\beta_{SH}\omega_{n,J}}\right)
\end{align}
We will fix the expectation value of the energy to be the mass $M$ of a large BTZ black hole. We will also set the cut-off in $J$ to be $J_{cut}$ and fix it by matching with the Bekenstein-Hawking entropy. We will eventually see that it is self-consistent to work in the low-lying part of the spectrum where $\omega_{n,J} \approx -\frac{2nr_{h}\pi}{L^{2}\log(\frac{2\epsilon}{r_{h}})}$. This approximation is valid for $J$ such that $J \ll J_{max} = \frac{r_{h}^{3/2}}{L\sqrt{2\epsilon}}$. $J_{max}$ is where the expression for the low-lying part of the spectrum undergoes a complete breakdown. We will see that the matching of the entropy will lead to a $J_{cut}$ that is a few orders of magnitude smaller than $J_{max}$. This means that we are in the regime where the (quasi-)degeneracy of the low-lying normal modes can be exploited to great effect. 

Before proceeding, we fix $\epsilon$ by demanding that the proper invariant distance from the horizon to the stretched horizon is the Planck length $l_{p}$\footnote{As discussed in the Introduction, outside of a Planck length (or a string length) we expect bulk EFT to be a good enough approximation of the UV complete description. In fact, we will eventually see that it is only a specific combination of $J_{cut}$ and $\epsilon$ that ultimately affects our results. So there is some freedom in these choices.}. This also makes the stretched horizon, coordinate independent.  The proper distance $s$ from horizon to the cut-off surface is
\begin{align}
\frac{s}{L} = \int_{r_{h}}^{r_{h}+\epsilon}\frac{dr}{\sqrt{r^2-r_{h}^2}}
      =  \coth^{-1}\Bigg[\frac{r_{h}+\epsilon}{\sqrt{\epsilon(2r_{h}+\epsilon)}}\Bigg] \approx \frac{1}{2}\log\Bigg[\frac{r_{h}+\epsilon+\sqrt{2r_{h}\epsilon}}{r_{h}+\epsilon-\sqrt{2r_{h}\epsilon}}\Bigg]
\end{align}
Here $\approx$ denotes the approximation where we have ignored the $\epsilon^2$ term. After rearranging terms
and again ignoring $\epsilon^2$ terms, we end up with  
\begin{align}
    \frac{r_{h}}{2\epsilon} = \frac{4 e^{2s/L}}{(e^{2s/L}-1)^2}
\end{align}
As we mentioned, we need to fix the geodesic distance $s$ to be the Plank length $l_{p}$. Solving for $\epsilon$
for small $l_{p}$, we get 
\begin{align}
    \epsilon = \frac{r_{h}}{2}\frac{l_{p}^2}{L^2} = \frac{r_{h}}{2}\frac{G_{N}^{2}}{L^2}.
\end{align}
where we have used the fact that in three dimension, $l_{p} = G_{N}$. 
Note that with this choice, $J_{max}$ takes the simple form
$J_{max}=\frac{r_h}{l_p}$. 

For small $\epsilon$, which corresponds to small $G_N/L$,  we can write\footnote{Note that the continuum approximation in our analysis is related to the semi-classical gravity approximation $N \rightarrow \infty$.}
\begin{align}
\log Z_{SH} &\approx L^{2}\frac{\log(\frac{2\epsilon}{r_{h}})}{2\pi r_{h}}\int^{\infty}_{0}\sum_{J=-J_{cut}}^{J_{cut}}d\omega \log\left(1-e^{-\beta_{SH}\omega}\right) \nonumber \\
& = -L^{2}\log(\frac{2\epsilon} {r_{h}})\frac{\pi}{12r_{h}\beta_{SH}}(2J_{cut}+1) \approx -L^{2}\log(\frac{2\epsilon}{r_{h}})\frac{\pi}{6r_{h}\beta_{SH}}J_{cut}
\end{align}
We take $J_{cut}$ to be sufficiently large to make the last step valid, but it is easy to check that it can still be small enough to make the low-lying spectrum reliable. To compute the density of states, we use
\begin{align}
Z_{SH}\approx \exp\left(L^{2}\log(\frac{r_{h}}{2\epsilon})\frac{\pi}{6r_{h}\beta_{SH}}J_{cut}\right) \equiv \int dE \rho(E) e^{-\beta_{SH}E}
\end{align}
Where $\rho(E)$ is the density of state for a fixed energy $E \equiv \sum_{n,J}\omega_{n,J}N_{n,J}$. Thus one can obtain $\rho(E)$ as the inverse Laplace transform of the partition function
\begin{align}
\rho(E) =  \frac{1}{2\pi i}\int^{C+i\infty}_{C-i\infty}d\beta_{SH} \exp\left[L^{2}\log(\frac{r_{h}}{2\epsilon})\frac{\pi}{6r_{h}\beta_{SH}}J_{cut}+\beta_{SH}E\right]
\end{align} 
We can evaluate the above integral using a saddle point approximation. The saddle is at 
$\beta^{*}_{SH} = \sqrt{L^{2}\log(\frac{r_{h}}{2\epsilon})\frac{\pi}{6r_{h} E}J_{cut}}$, which leads to the following expression for the density of states:
\begin{align}\label{rho}
\rho(E) =  \exp\left[ 2\sqrt{L^{2}\frac{\pi\log(\frac{r_{h}}{2\epsilon}) E}{6r_{h}}J_{cut}}\right]
\end{align}
We take $E$ to be the mass $M$ of the black hole. This essentially specifies the statistical ensemble. By taking log of the density of states we get the microcanonical entropy
\begin{align}
S_{SH}(M) =  2\sqrt{L^{2}\frac{\pi\log(\frac{r_{h}}{2\epsilon}) M}{6r_{h}}J_{cut}}= 2\pi\sqrt{\frac{M L^{2}}{2G_{N}}\Bigg[\frac{2G_{N}\log\big(\frac{r_{h}}{2\epsilon}\big)J_{cut}}{6\pi r_{h}}\Bigg]} 
\end{align}
Demanding the entropy to be the entropy associated to a BTZ black hole of mass $M=E (= \frac{r_{h}^{2}}{8G_{N}L^{2}})$, we have\footnote{Note that it is a non-trivial fact that the entropy is proportional to $\sqrt{E}$. We would not be able to get a match (no matter what the $J_{cut}$) if this were not the case.}
\begin{align}
     S_{BTZ}(E)= 2\pi\sqrt{\frac{EL^{2}}{2G_{N}}} = \frac{\pi r_{h}}{2G_{N}}
\end{align}
Equating the two expressions for the entropy fixes $J_{cut}$ to be 
\begin{align}\label{j cut}
    J_{cut} = \frac{3\pi r_{h}^{3/2}}{\sqrt{2\epsilon}L\log\big(\frac{r_{h}}{2\epsilon}\big)}=\frac{3\pi}{\log\frac{r_{h}}{2\epsilon}} J_{max}
\end{align}
From the last expression, one can easily see that $J_{cut} \ll J_{max}$ because of the hierarchy between the horizon radius and the Planck length. This is not a huge hierarchy, but it {\em is} a hierarchy. It is roughly the log of the ratio between the horizon radius and Planck length and in the real world that is a factor of about $\sim 100$. Remarkably, one can show that a factor $\gtrsim 10$ is eminently sufficient to bring us squarely into the regime of the spectrum that is responsible for the linear ramp in the spectral form factor expected in quantum chaotic systems \cite{Pradipta} (and as we will see in this paper, thermal aspects of black holes).

Remarkably, (\ref{j cut}) leads to identifying $\beta^{*}_{SH}$ to be {\em precisely} the Hawking temperature $\beta_{H}$:
\begin{align}\label{hawking temp}
   \beta_{SH}^{*} = \frac{2\pi L^{2}}{r_{h}} = \beta_{H}
\end{align}
Note that the fixing (\ref{j cut}) simplifies the expression of $\rho(E)$ in (\ref{rho})
\begin{align}
  \rho(E) =  \exp\left[ 2\pi\sqrt{\frac{E L^2}{2G_{N}}}\right]  
\end{align}
This is the Cardy density of states \cite{Strominger}\footnote{This counts both the left and right moving sectors together. In fact, a version of the calculation we presented here, that can reproduce the full holomorphically factorized Cardy formula, also exists \cite{Pradipta}.}. We will make more comments about connections to the dual CFT, elsewhere. 


A key observation in this calculation was that the $J_{cut}$ was determined in terms of $\epsilon$. This will turn out to be a very convenient feature in the ``continuum" limit, in reproducing the Hartle-Hawking correlator in a later section. The results of this section will play an implicit role there, in fixing the temperature and entropy correctly. But the HH correlator requires more -- a precise cancellation between the Klein-Gordon normalization and the $\log (r_h/\epsilon)$ in the normal modes will also turn out to be crucial.

\section{Vacuum, Excited States and Correlators}

\noindent The mode solutions of the scalar field equation can be written in the form 
\begin{align}\label{Mode}
\mathcal{U}_{n,J}(r,t,\psi) = \frac{1}{\sqrt{r}}e^{-i\omega_{n,J} t}e^{iJ\psi}\phi_{{n},J}(r) 
\end{align}
We have written $\phi_{n,J}$ instead of $\phi_{\omega,J}$ to emphasize that the continuous spectrum of $\omega$'s becomes quantized when we put a Dirichlet boundary condition at the stretched horizon. We can write the general field solution $\Phi$ in the form of a mode expansion 
\begin{align}\label{Mode expansion}
 \Phi(r,t,\psi) = \sum_{n,J}\frac{1}{\sqrt{4\pi\omega_{n,J}}}\left(b_{n,J}\mathcal{U}_{n,J}(r,t,\psi)+b^{\dagger}_{n,J}\mathcal{U}^{*}_{n,J}(r,t,\psi)\right) 
\end{align}
We will use the Klein-Gordon inner product to normalize the modes $\mathcal{U}_{n,J}$. This is done in Appendix \ref{KGApp}. $b_{n,J}$ and $b^{\dagger}_{n,J}$ are annihilation and creation operators for the modes. The stretched horizon vacuum is defined as
\begin{align}
    b_{{n,J}}|0\rangle_{SH} = 0 \hspace{0.5cm} \forall \hspace{0.3cm} n \in\mathbb{N}, \  J \in \pm \mathbb{Z}^+
\end{align}
Note that this has similarities to the Boulware vacuum -- but a crucial difference is that the modes are defined to vanish at the stretched horizon, instead of being defined all the way to horizon\footnote{See \cite{Emparan} for a recent discussion of the Boulware state in holography, for (near-)extremal black holes.}. We will define $\Tilde{\mathcal{U}}_{n,J} =\frac{1}{\sqrt{4\pi\omega_{nJ}}}\mathcal{U}_{n,J}$ for convenience. 

A crucial point is that we are not simply interested in correlators on the above vacuum. Motivated by the calculations in the previous section, we will be interested in computing correlators in a typical excited state built on the above vacuum, with energy equal to that of the black hole mass. Such excited states can be constructed as 
\begin{align}\label{excited state}
   |N\rangle\equiv |\{N_{mK}\}\rangle = \prod_{m,K}\frac{1}{\sqrt{N_{mK}!}}(b^{\dagger}_{mK})^{N_{mK}}|0\rangle_{SH}
\end{align}
We will eventually connect excitation energy of the level to the black hole mass, see \eqref{total energy}. Note that in making this identification, we are exploiting the fact that the energy measured by the exterior Schwarzschild time coordinate is the boundary (ADM) energy. 

In this state, the bulk Wightman function, defined as
\bea
\mathcal{G}_{c}^{+}\equiv\langle\{N_{mK}\}|\Phi(r,t,\psi)\Phi(r',t',\psi')|\{N_{mK}\}\rangle
\eea
becomes
\begin{align}\label{G}
&\mathcal{G}_{c}^{+}=\sum_{n,J,n',J'}\Big\langle\{N_{mK}\}|\big[b_{nJ}\Tilde{\mathcal{U}}_{nJ}(r)+b^{\dagger}_{nJ}\Tilde{\mathcal{U}}_{nJ}^{*}(r)\big]\big[b_{n'J'}\Tilde{\mathcal{U}}_{n'J'}(r')+b^{\dagger}_{n'J'}\Tilde{\mathcal{U}}_{n'J'}^{*}(r')\big]|\{N_{mK}\}\Big\rangle  \nonumber \\
& \hspace{5mm}= \sum_{n,J,n',J'}\Big\langle\{N_{mK}\}\Big|\Big[b_{nJ}b_{n'J'}\Tilde{\mathcal{U}}_{nJ}(r)\Tilde{\mathcal{U}}_{n'J'}(r')+b_{nJ}b^{\dagger}_{n'J'}\Tilde{\mathcal{U}}_{nJ}(r)\Tilde{\mathcal{U}}^{*}_{n'J'}(r')+ \nonumber \\ &\hspace{4cm}+b^{\dagger}_{nJ}b_{n'J'}\Tilde{\mathcal{U}}^{*}_{nJ}(r)\Tilde{\mathcal{U}}_{n'J'}(r')+b^{\dagger}_{nJ}b^{\dagger}_{n'J'}\Tilde{\mathcal{U}}^{*}_{nJ}(r)\Tilde{\mathcal{U}}^{*}_{n'J'}(r')\Big]\Big|\{N_{mK}\}\Big\rangle
\end{align}
\noindent The creation and annihilation operators act on such states in the following way
\begin{align}\label{annihlation}
  \hspace{-10mm}  b_{n,J}|N\rangle = b_{n,J}|N_{n1,J1},N_{n2,J2},....,N_{n,J},....\rangle = \sqrt{N_{n,J}}|N_{n1,J1},N_{n2,J2},....,N_{n,J}-1,....\rangle 
\end{align}
\begin{align}\label{creation}
     \hspace{-4.5mm}b_{n,J}^{\dagger}|N\rangle = b_{n,J}|N_{n1,J1},N_{n2,J2},....,N_{n,J},....\rangle = \sqrt{N_{n,J}+1}|N_{n1,J1},N_{n2,J2},....,N_{n,J}+1,....\rangle
\end{align}
Using (\ref{annihlation}) and (\ref{creation}) in equation (\ref{G}), we see that only the second and third terms survive. 
Now, using the following identities in equation (\ref{G})
\begin{align}\label{ac}
    \langle \{N_{ml}\}|b_{nJ}b^{\dagger}_{n'J'}|\{N_{ml}\} \rangle = \sqrt{N_{n'J'}+1}\sqrt{N_{nJ}+1}\delta_{n,n'}\delta_{J,J'}
\end{align}
\begin{align}\label{ca}
    \hspace{-14mm}\langle \{N_{ml}\}|b^{\dagger}_{nJ}b_{n'J'}|\{N_{ml}\} \rangle = \sqrt{N_{n'J'}}\sqrt{N_{nJ}}\delta_{n,n'}\delta_{J,J'}
\end{align}
together with equations (\ref{annihlation}), (\ref{creation}), (\ref{ac}), and (\ref{ca}) into (\ref{G}), we finally get
\begin{align}\label{bulk Wightman1}
    \mathcal{G}_{c}^{+} = \sum_{n,J}\Big[ (N_{nJ}+1)\ \Tilde{\mathcal{U}}_{nJ}(r,t,\psi)\ \Tilde{\mathcal{U}}^{*}_{nJ}(r',t',\psi')+N_{nJ}\ \Tilde{\mathcal{U}}^{*}_{nJ}(r,t,\psi)\ \Tilde{\mathcal{U}}_{nJ}(r',t',\psi') \Big]
\end{align} 
Putting $\Tilde{\mathcal{U}}_{n,J} = \frac{1}{\sqrt{4\pi\omega_{n,J}}}\mathcal{U}_{n,J}$ in (\ref{bulk Wightman1}) and using the reality condition of $\phi_{n,J}(r)=\phi_{n,J}^{*}(r)$, this becomes
\begin{align}\label{bulk Wightman2}
&\mathcal{G}^{+}_{c} = \sum_{n,J}\frac{1}{4\pi\omega_{nJ}}\frac{1}{\sqrt{rr'}}\Big[(N_{nJ}+1)e^{-i\omega_{nJ}(t-t')}e^{iJ(\psi-\psi')}+N_{nJ}e^{i\omega_{nJ}(t-t')}e^{-iJ(\psi-\psi')}\Big] \times \\
&\hspace{10cm}\times\phi_{nJ}^{*}(r)\phi_{nJ}(r')\nonumber
\end{align}

We will be working with boundary correlators and we get them by taking the boundary limit $r,r'\rightarrow \infty$ of the above expression:
\begin{align}
G^{+}_{c} = \lim_{r,r'\rightarrow\infty}2\nu \Big(\frac{r}{L}\Big)^{\Delta}\ 2\nu \Big(\frac{r'}{L}\Big)^{\Delta}\ \mathcal{G}^{+}_{c}
\end{align}
More explicitly, this leads to
\begin{align}\label{cut-off two-point function}
&\hspace{-6mm}G^{+}_{c}=\lim_{r,r'\rightarrow\infty}\frac{r^{\Delta}r'^{\Delta}}{L^{2 \Delta}\sqrt{rr'}} \sum_{n,J}\frac{(2\nu)^{2}}{4\pi\omega_{nJ}}\Big[ (N_{nJ}+1)e^{-i\omega_{nJ}(t-t')}e^{iJ(\psi-\psi')} + N_{nJ}e^{i\omega_{nJ}(t-t')}e^{-iJ(\psi-\psi')}\Big]\times\\
&\hspace{10cm}\times\phi_{nJ}^{*}(r)\phi_{nJ}(r')\nonumber
\end{align}
\noindent This is the boundary Wightman function in the eigenstate $|N_{mK}\rangle$. The field $\phi_{n,J}(r)$ is not normalized yet, we will normalize it in the next subsection and Appendix \ref{NormApp}. 

Let us also present the Wightman function on the stretched horizon vacuum for completeness. It is given as $_{SH}\langle0|\Phi(r,t,\psi)\Phi(r',t'\psi')|0\rangle_{SH}$ and takes the form
\begin{align}\label{vacuum G}
&\Tilde{\mathcal{G}}^{+}_{c}\equiv {}_{SH}\langle0|\Phi(r,t,\psi)\Phi(r',t'\psi')|0\rangle_{SH}\nonumber \\
&\hspace{5mm}= \sum_{n,J}\frac{1}{4\pi\omega_{n,J}}
\frac{1}{\sqrt{rr'}}\phi_{n,J}^{*}(r)\phi_{n,J}(r')e^{-i\omega_{n,J}(t-t')}e^{iJ(\psi-\psi')} 
\end{align}
We get the boundary Wightman function by again taking the limit $r,r'\rightarrow \infty$.
\begin{align}
\Tilde{G}^{+}_{c} = \lim_{r,r'\rightarrow\infty}2\nu \Big(\frac{r}{L}\Big)^{\Delta}\ 2\nu \Big(\frac{r'}{L}\Big)^{\Delta}\ \Tilde{\mathcal{G}}^{+}_{c}
\end{align}
This takes the form
\begin{align}\label{vacuum boundary G}
\Tilde{G}^{+}_{c} = \lim_{r,r'\rightarrow\infty}\frac{r^{\Delta}r'^{\Delta}}{L^{2 \Delta}\sqrt{rr'}}\sum_{n,J}\frac{(2\nu)^2}{4\pi\omega_{n,J}}\phi_{n,J}^{*}(r)\phi_{n,J}(r')e^{-i\omega_{n,J}(t-t')}e^{iJ(\psi-\psi')}
\end{align}
This is the boundary Wightman function in the stretched horizon vacuum. Note that this is a very explicit form, because we have complete expressions for the radial modes including their normalizations (which we discuss momentarily).
 
\subsection{Normalization}

The Klein-Gordon inner product is discussed in Appendix \ref{KGApp}, and
for the stretched horizon BTZ it leads to the following equation for the radial solutions:
\begin{align}\label{KG}
\int^{\infty}_{r_{h}+\epsilon}dr \frac{L^2}{r^{2}-r_{h}^{2}}\phi_{n,J}\phi^{*}_{n',J} = \delta_{n,n'}
\end{align}
Here $\epsilon$ is the location of the stretched horizon which is of the order of the Planck length ($\epsilon\ll r_{h}$) as noted in the previous section. The normalizability of the modes near the boundary leads us to the following expression\footnote{Note that here we denote $\frac{\omega_{n,J}L^{2}}{r_{h}}\rightarrow \omega$ and $\frac{JL}{r_{h}}\rightarrow J$. Also, we are writing $\omega_{n,J}$ to be $\omega$ for convenience. So here $\omega$ should be understood as quantized modes.}, see equation (\ref{normalized at boundary}):
\begin{align}\label{phi1}
&\phi_{n,J}(r) = C_{n,J} S r^{\frac{1}{2}-iJ} (r^{2}-r_{h}^{2})^{-\frac{i\omega}{2}}e^{-\frac{i\pi}{2}(1-i\omega-iJ+\nu)} \left(\frac{r}{r_{h}}\right)^{-(1-i\omega-iJ+\nu)} 
 {}_2F_{1}\left(\alpha,\beta;\Delta;y\right)
\end{align}
Here $C_{n,J}$ is an undetermined constant which we would like to fix using (\ref{KG}). However, from (\ref{KG}) one can only fix $|C_{n,J}|$. To fix the ${\rm Arg}(C_{n,J})$, we will use the reality condition of the modes, i.e.
\begin{align}\label{reality}
\phi_{n,J}(r)=\phi^{*}_{n,J}(r)
\end{align}
The calculation of the normalization factor $C_{n,J}$ is given in Appendix \ref{NormApp}. Here, we will just present the result:
\begin{align}\label{normalization}
    C_{n,J} = \frac{1}{L \sqrt{\log\Big(\frac{r_{h}}{2\epsilon}\Big)}}\Big(1-\frac{\epsilon}{r_{h}}\Big)^{-1-\nu}r_{h}^{\frac{iL}{r_{h}}(\omega L+J)}e^{i\pi\beta}\Bigg[\frac{\Gamma(\beta)\Gamma(\alpha^{*})\Gamma\Big(\frac{iJL}{r_{h}}\Big)^{2}}{\Gamma(\beta^{*})\Gamma(\alpha)\Gamma\Big(\frac{i\omega L^{2}}{r_{h}}\Big)\Gamma\Big(-\frac{i\omega L^{2}}{r_{h}}\Big)}\Bigg]^{\frac{1}{2}}
\end{align}
 Using (\ref{normalization}) the expression given in (\ref{phi1}) becomes
\begin{align}\label{phi final}
&\phi_{n,J}(r) = \frac{T}{L \sqrt{\log(\frac{r_{h}}{2\epsilon})}}r^{-\frac{1}{2}-\nu}\frac{r_{h}^{1+\nu}}{\Gamma(\nu+1)}\left(1-\frac{\epsilon}{r_{h}}\right)^{-1-\nu}\left(1-\frac{r_{h}^{2}}{r^{2}}\right)^{-\frac{i\omega}{2}} {}_{2}F_{1}\left(\alpha,\beta,\Delta,y\right)
\end{align}
where 
\bea
T = \left[\frac{\Gamma(\beta^{*})\Gamma(\alpha)\Gamma(\beta)\Gamma(\alpha^{*})}{\Gamma(\frac{i\omega L^{2}}{r_{h}})\Gamma(-\frac{i\omega L^{2}}{r_{h}})}\right]^{\frac{1}{2}}.
\eea

\section{Emergence of the Smooth Horizon Correlator}\label{HHSection}

We have found the normalization $C_{n,J}$ in Section 4 in equation (\ref{normalization}). The normalized field is given in equation (\ref{phi final}). Now we can use equation (\ref{cut-off two-point function}) to write the explicit expression for the boundary Wightman function for the stretched horizon case. Equation (\ref{cut-off two-point function}) becomes
\begin{align}\label{cutoff correlator expression}
&\hspace{-0.5cm}G^{+}_{c} = \frac{K r_{h}^{2\Delta}}{(\Gamma(\Delta))^2}\sum_{n,J}\frac{(2\nu)^{2}}{4\pi\omega_{nJ}}\Big[(N_{nJ}+1) e^{-i\omega_{n,J}(t-t')}e^{iJ(\psi-\psi')}+N_{nJ} e^{i\omega_{n,J}(t-t')}e^{-iJ(\psi-\psi')}\Big]\times\\
&\hspace{12.5cm}\times\frac{|f(\omega_{n,J},J)|^{2}}{L^{2 \Delta +2}} \nonumber
\end{align}
where, 
\bea
f(\omega_{n,J},J)\equiv\frac{\Gamma(\beta^{*})\Gamma(\alpha)}{\Gamma\big(\frac{i\omega_{n,J}L^2}{r_{h}}\big)}, \ \ {\rm and} \ \ K\equiv \frac{1}{\log(\frac{r_{h}}{2\epsilon})}\left(1-\frac{\epsilon}{r_{h}}\right)^{-2\Delta}.
\eea 
For the massless scalar field, we use $\nu = 1$. The goal of this section is to study this object and discuss the emergence of the Hartle-Hawking correlator from it, as well as the finite-$N$ corrections.

We start with the observation that $|f(\omega_{n,J},J)|^{2}$ can be simplified further by using the gamma function identities,
\begin{align}\label{gamma identity 1}
\Gamma(1+ik) = ik\Gamma(ik)
\end{align}
\begin{align}\label{gamma identity 2}
\Gamma(ik)\Gamma(-ik) = \frac{\pi}{k\sinh(\pi k)}
\end{align}
Using (\ref{gamma identity 1}) and (\ref{gamma identity 2}) into (\ref{cutoff correlator expression}), $f(\omega_{n,J},J)$ simplifies to
\begin{align}
    |f(\omega_{n,J},J)|^{2} = \frac{\pi L^4]}{r_{h}^{3}} \frac{\omega_{n,J}\left(\frac{(\omega_{n,J}^2 L^{2}-J^2)}{4}\right)\sinh[\frac{\pi\omega_{n,J}L^{2}}{r_{h}}]}{\sinh[\frac{\pi L(\omega_{n,J}L+J)}{2r_{h}}]\sinh[\frac{\pi L(\omega_{n,J}L-J)}{2r_{h}}]}\equiv\frac{\pi \omega_{n,J}L^4}{r_{h}^{3}}\Tilde{f}(\omega_{n,J},J)
\end{align}
For $\nu=1$, $\Delta = 2$ and $r_{h}^{2\Delta} = r_{h}^{4}$. A factor of $\frac{1}{r_{h}^{3}}$ is coming from $|f(\omega_{n,J},J)|^{2}$, so in the expression of $G_{c}^{+}$ only one factor of $r_{h}$ remains. Here we are emphasizing the dependence of $r_{h}$ as a prefactor of $G_{c}^{+}$ because the final expression of this boundary correlator will be independent of $r_{h}$. From (\ref{cutoff correlator expression})
\begin{align}\label{G sum}
&\hspace{-1cm}L^2G^{+}_{c} = r_{h} K \sum_{n,J}N_{nJ}\Bigg[e^{-i\omega_{n,J}(t-t')}e^{iJ(\psi-\psi')} + e^{i\omega_{n,J}(t-t')}e^{-iJ(\psi-\psi')}\Bigg]\Tilde{f}(\omega_{n,J},J)+ \\
&\hspace{7.5cm}+ r_{h} K\sum_{n,J}\Tilde{f}(\omega_{n,J},J) e^{-i\omega_{n,J}(t-t')}e^{iJ(\psi-\psi')}\nonumber
\end{align}

As we mentioned before, we are considering the states lying in the energy level at the mass of the black hole. 
\begin{align}\label{total energy}
    E = M = \sum_{n,J}\omega_{n,J}N_{nJ}
\end{align}
For large $E$, we expect that it should be possible to replace the occupation number $N_{n,J}$ with it's mean value $\langle {N}_{nJ} \rangle$ i.e. $N_{nJ} \approx \langle {N}_{nJ} \rangle$. However, this is only possible if the variance of such number operators are suppressed sufficiently -- we will demonstrate this in the next subsection. From standard statistical analysis, if a level $N_{level}$ is integer partitioned as $\sum_{n}nN_{n}=N_{level}$ then the partitions $N_{n}$ are distributed according to the Boltzmann distribution with a Bose-Einstein mean\footnote{For a proof, see Appendix B of \cite{Bala}.}
\begin{align}\label{mean}
    \langle N_{n} \rangle =\frac{1}{e^{\pi n/\sqrt{6 N_{level}}}-1}
\end{align}
We can use this fact here in our analysis\footnote{To study typicality in 2d CFT, this fact has been used in \cite{Datta, Datta2}. We thank Shouvik Datta for collaborations on some related topics.}. Since $\omega_{n,J} = \frac{2n\pi r_{h}}{L^{2}\log(\frac{r_{h}}{2\epsilon})}$, we can write  
\bea
E = \frac{2\pi r_{h}}{L^{2}\log(\frac{r_{h}}{2\epsilon})}\sum_{n,J}n N_{nJ}
\eea
Assuming $N_{nJ}$ to be independent of $J$ \footnote{This is well supported by the fact that we are working in a limit where $\omega_{nJ}$ has no explicit $J$ dependence. Hence we expect the occupation number is also independent of $J$.} and writing $\sum_{J=-J_{cut}}^{J = J_{cut}} = 2J_{cut}+1 \approx 2J_{cut}$ for large $J_{cut} \ll J_{max}$ and using (\ref{j cut}) and $E=M$ in (\ref{total energy}) we finally obtain
\bea\label{micro us}
 \sum_{n}nN_{n} = \frac{\left(\log\big(\frac{r_{h}}{2\epsilon}\big)\right)^{2}}{96\pi^{2}}\equiv N_{level}
\eea
Comparing (\ref{micro us}) with (\ref{mean}) the mean occupation number is
\begin{align}
    \langle N_{n}\rangle = \frac{1}{e^{\frac{4\pi^{2}n}{\log\big(\frac{r_{h}}{2\epsilon}\big)}}-1}
\end{align}
Since $\beta\omega_{n,J} = \frac{2\pi L^{2}}{r_{h}}\frac{2n\pi r_{h}}{L^{2}\log\big(\frac{r_{h}}{2\epsilon}\big)} = \frac{4\pi^2 n}{\log\big(\frac{r_{h}}{2\epsilon}\big)}$, this reduces precisely to
\begin{align}
    \langle N_{n} \rangle = \frac{1}{e^{\beta\omega_{n,J}}-1}.
\end{align}
where $\beta$ is the Hawking temperature.

The above result means that setting $N_{n} = \langle N_{n} \rangle$ and plugging it back into equation (\ref{G sum}), we get
\begin{align}\label{G approx}
&\hspace{-1cm}G^{+}_{c} = r_{h}K\sum_{n=0}^{\infty}\sum_{J=-J_{cut}}^{J_{cut}}\frac{1}{e^{\beta\omega_{n,J}}-1} \left[e^{-i\omega_{n,J}(t-t')}e^{iJ(\psi-\psi')}+e^{i\omega_{n,J}(t-t')}e^{-iJ(\psi-\psi')}\right]\Tilde{f}(\omega_{n,J},J)+\\
& \hspace{6.5cm}+r_{h}K\sum_{n=0}^{\infty}\sum_{J=-J_{cut}}^{J_{cut}}\Tilde{f}(\omega_{n,J},J)e^{-i\omega_{n,J}(t-t')}e^{iJ(\psi-\psi')}\nonumber
\end{align}
Using the identification $\Tilde{f}(-\omega,-J)=-\Tilde{f}(\omega,J)$, we can easily check that (\ref{G approx}) simplifies to
\begin{align}\label{SH 2pt}
     G^{+}_{c} = r_{h}K\sum_{n=-\infty}^{\infty}\sum_{J=-J_{cut}}^{J_{cut}}\frac{e^{\beta\omega_{n,J}}}{e^{\beta\omega_{n,J}}-1} \left[\Tilde{f}(\omega_{n,J},J)e^{-i\omega_{n,J}(t-t')}e^{iJ(\psi-\psi')}\right]
\end{align}

We can now consider the limit where $\epsilon$ is sent to zero. This corresponds to the large-$N$ limit (or the $G_N \rightarrow 0$ limit). This means that using the fact that $\omega_{n+1,J}-\omega_{n,J}=\frac{2\pi r_{h}}{\log(\frac{r_{h}}{2\epsilon})}$, we can convert the sum to an integral. Crucially, we also have $r_{h}K \approx \frac{r_{h}}{\log\Big(\frac{r_{h}}{2\epsilon}\Big)}$. Quite remarkably, while converting the sum into integral, the $r_{h}K$ factor goes away completely due to the $r_{h}$ and $\log\Big(\frac{2\epsilon}{r_{h}}\Big)$ dependence of the normal modes as in (\ref{normal modes}). Equally noteworthy is the fact that the $J_{cut}$ in \eqref{j cut} is such that $J_{cut} \rightarrow \infty$ in the same limit. The end result is that in this limit, we have
\begin{equation}\label{G HH}
G^{+}_{c} \rightarrow G^{+}_{HH} = \int_{-\infty}^{\infty}\frac{d\omega}{2\pi}\sum_{J=-\infty}^{\infty}\frac{e^{\beta\omega}}{e^{\beta\omega}-1}\Tilde{f}(\omega,J)e^{-i\omega(t-t')}e^{iJ(\psi-\psi')}
\end{equation}
which is exactly the boundary Hartle-Hawking correlator, written as a Fourier transform from momentum space. This form can be found, eg. in \cite{Festuccia, Yang, Zhiboedov}. Here
\bea
\Tilde{f}(\omega,J) \equiv \frac{\frac{(\omega^2 L^{2}-J^2)}{4}\sinh\big[\frac{\pi\omega L^{2}}{r_{h}}\big]}{\sinh\big[\frac{\pi L(\omega L+J)}{2r_{h}}\big]\sinh\big[\frac{\pi L(\omega L-J)}{2r_{h}}\big]}.
\eea
which can be re-written in terms of Gamma functions in familiar forms. Note that the momentum space version of the two-point function (\ref{G HH}) satisfies the KMS condition 
\begin{align}
    G^{+}_{c}(-\omega,J) = e^{-\beta\omega}G^{+}_{c}(\omega,J)
\end{align}

The Fourier transform above can in fact be explicitly done \cite{KrishnanRotation}, and one can show that it leads to the position space Hartle-Hawking correlator. The latter is known to be the result that one obtains from the covering space AdS$_3$ Green function when one constructs BTZ Green function as a quotient of AdS$_3$ via a mirror sum \cite{Ichinose, Lifschytz}. See eg. eqn (2.5) in \cite{Maldacena} for a convenient explicit expression. This position space correlator with $\psi=\psi'$ is useful for discussing the temporal decay of the correlator. 

What we have shown above is that in the $\epsilon$ (or $G_{N}$ or $1/N$) $\rightarrow 0$ limit, $J_{cut} \rightarrow \infty$ and hence $G^{+}_{c}\rightarrow G^{+}_{HH}$. In the $t \rightarrow \infty$ limit, this correlator decays exponentially to zero. This is a version of the information paradox, and to resolve it, we need a contribution that kicks in at timescales of $\mathcal{O}(S_{BH})$. This is the analogue of the Page time for this problem. We will see in the next subsection that in our calculation, the stretched horizon automatically saves unitarity -- we get another contribution from the suppressed variance in the typical states which will not decay in this limit.

\subsection{Exponential Suppression of Variance}\label{sup-var}

In proving the equivalence between Hartle-Hawking two point function (\ref{G HH}) and the large-$N$ limit of the excited state stretched horizon two point function (\ref{cutoff correlator expression}), a vital step we used was the identification of $N_{n} = \Bar{N}_{n} \equiv \langle N_{n} \rangle$. We have seen in our  state at large fixed $E$, that the mean occupation number $\langle{N}_{n}\rangle$ is that of the canonical ensemble with fixed temperature $\beta$. However the replacement $N_{n} = \Bar{N}_{n}$ is only valid when the variance to mean ratio of the term associated to $N_{n}$ in the expression of the two point function is sufficiently suppressed.  The key observation of this subsection is that this indeed happens for small but finite $\epsilon$ for typical states at a given energy (or a microcanonical sliver). Note that our previous calculation that matched the Hartle-Hawking correlator was done directly with eigenstates -- but due to the suppressed variance, it also applies to typical states.

Our argument follows directly from previous works \cite{Lloyd, Bala, Datta}. We partially review this in the language of \cite{Bala} below to give a flavor of the ideas, and then make some comments specific to our case. Consider a state $|\psi\rangle$ as a superposition of energy eigenstate $|N\rangle$ having energy $(E,E+dE)$ in a small energy window of $dE$.
\begin{align}
    |\psi\rangle \equiv \sum_{N}c^{\psi}_{N}|N\rangle = \sum_{\{N_{n}\}} c^{\psi}_{N}|\{N_{n}\}\rangle; \; \hspace{1cm}\text{s.t.}\hspace{0.5cm} \; \sum_{n,J}\omega_{n,J}N_{n} = E.
\end{align}
We are considering an ensemble of states $|\psi\rangle$ denoted as $M_{\psi}$ such that the energy eigenvalues of the eigenstates constituting $|\psi\rangle$ lie in the energy sliver $E$ and $E+dE$. The dimension of the Hilbert space constructed out of those restricted states is $\sim e^{S(E)}$, where $S(E)$ is the entropy of the ensemble of states with energy $E$ -- there are $\sim e^{S(E)}$ eigenstates in this sliver. In any such state, the expectation value of an operator is given by:
\bea
    \langle \psi| \mathcal{O} |\psi \rangle = \sum_{N, N'} c^{\psi}_{N} c^{\psi *}_{N'} \langle N'|\mathcal{O}|N\rangle.
\eea
Here $\sum_{N} |c^{\psi} _{N}|^{2} = 1$.
The above formula is just the conventional quantum expectation value on a general state in the microcanonical sliver Hilbert space. 

Now the (microscanonical) ensemble average of the expectation value of $\mathcal{O}$ over all the pure states in the sliver Hilbert space $M_\psi$ can be defined via
\begin{align}\label{demand 1}
 \langle \mathcal{O} \rangle= \int D\psi    \langle \psi| \mathcal{O} |\psi \rangle = \int \vec{dc^{\psi}} \sum_{N, N'} c^{\psi}_{N} c^{\psi *}_{N'} \langle N'|\mathcal{O}|N\rangle
\end{align}
where the measure $\int D\psi \equiv \int \vec{dc^{\psi}}$. We say that this is what one expects as a result of a $typical$ measurement of $\mathcal{O}$ in the sliver.  Note that the goal of the ensemble average here is only to define what one means by a typical state and characterize the measurements of $\mathcal{O}$ on them. Now, this measure over the Hilbert space satisfies \cite{Bala}
\begin{align}\label{useful id1}
    \int \vec{dc^{\psi}} |c^{\psi} _{N}|^{2} = \frac{1}{e^{S}}.
\end{align}
Using this, as well as $\sum_{N} |c^{\psi} _{N}|^{2} = 1$ and the normalization $\int \vec{dc^{\psi}}=1$ to evaluate the integral \eqref{demand 1}, we find that the ensemble average is simply the average over eigenstates\footnote{This is precisely the derivation of eqn (2.28) from (2.25) in \cite{Bala}, with the only difference that we are working with a single operator instead of a generating function.}:
\begin{align}\label{demand 2}
     \langle \mathcal{O} \rangle = e^{-S} \sum_{N} \mathcal{O}_{N} \equiv \langle \mathcal{O} \rangle_\mu, \; \text{where} \; \mathcal{O}_{N} \equiv \langle N|\mathcal{O}|N\rangle.
 \end{align}
The $\langle \rangle_\mu$ stands for the average over eigenstates. The average $\langle \rangle$ on the LHS is what we will call the microcanonical average (and up to exponentially suppressed variance corrections, the result will be the same in the canonical ensemble as well, as we know from elementary statistical mechanics). What the above result shows is that the microcanonical ensemble average is simply equal to the average over the eigenstates. For us expectation values of $\mathcal{O}$ would correspond to the two point functions that we have considered in the previous (sub-)sections. The above result is basically saying that 
the average expectation value of the operator on the ensemble of all microstates is the same as that in the ensemble of just the eigenstates.

Now the crucial statement is that when one is computing variances, something interesting happens. The precise details are unimportant for us (but see section 2.2 of \cite{Bala}). The interesting fact however is that the variance of operators in the ensemble of all microstates is {\em suppressed} by a factor of $e^{-S}$ relative to the variance in the ensemble of eigenstates. In other words
\bea
\delta \mathcal{O}^2 \equiv \langle \mathcal{O}^2 \rangle - \langle \mathcal{O} \rangle \langle \mathcal{O} \rangle \approx e^{-S} \left(\langle \mathcal{O}^2 \rangle_\mu - \langle \mathcal{O} \rangle_\mu \langle \mathcal{O} \rangle_\mu\right) \equiv e^{-S} \delta \mathcal{O}^2_\mu
\eea

What does this mean for our calculation of the stretched horizon correlator that reduced to the Hartle-Hawking correlator at small-$G_N$, ie., small-$\epsilon$? The key point is that we did our calculation in an excited eigenstate at some high energy level. The microcanonical sliver can essentially be viewed as this level, because there are a very large number of ways to populate this level\footnote{We could also consider a small sliver of levels, but it is not strictly necessary is our point. Note that if we include the weak $J$-dependence of the spectrum, it is more natural to work with a legitimate sliver.}. In any event, in this sliver we replaced $N_n$ with $\bar N_n$, with the latter turning out to be the canonical ensemble result. The crux of our observations above is that if we were to consider a typical state, this replacement is valid up to (square root of) variance corrections that are suppressed by $e^{-S/2}$. The $\mathcal{O}$ in our case corresponds to the the scalar two-point operator. In other words, the canonical replacement is an extremely good approximation on a typical state. This justifies our derivation of the HH correlator on typical states. 

Even though exponentially suppressed, this correction is of course crucial for resolving Maldacena's information paradox \cite{Maldacena}. We know from the form of the HH correlator that it decays exponentially with time, loosely as $ e^{-\pi t/\beta}$ as a result of the $\cosh \pi t/\beta$ terms in the denominator. This means that at timescales of the order of $\beta S_{BH}$ we expect that this expression would go below $e^{-S_{BH}}$ and we can no longer ignore the variance term. On a typical microstate, we should then no longer trust the HH result, and the behavior is controlled by the fluctuations. Note that this argument uses variance to argue about the size of the fluctuations in a typical pure state -- we are not merely making a statement about averaged quantities. Preliminary\footnote{We hope to report on a more detailed (numerical) study of the finite-time, finite-$G_N$ behavior of correlators in upcoming work. We suspect that including the $J$-dependence of the spectrum may allow us to study the physics of scrambling.} numerical experiments with specific values for the parameters indicates that this expectation about the behavior of the correlators, fluctuations and timescales is consistent with actual results.

If we take the limit $\epsilon \rightarrow 0$ or $G_{N} \rightarrow 0$ first, then due to infinite entropy the suppression due to variance correction goes away and the HH correlator becomes exact. In other words, Maldacena's information paradox is sharp if we $G_N\rightarrow 0$ first. At finite $N$ on the other hand, we can take $t \rightarrow \infty$ limit first and there is no paradox.

\section{Scrambling Time, Page Time and Typicality Time}

In the previous section, we have shown that the effective correlators of the stretched horizon are precisely those of a smooth horizon. It is useful to clarify some aspects of this statement, if we hope to understand the status of firewalls using these methods. Let us emphasize that the existence (or not) of firewalls is very much tied to the timescales under consideration. The original firewall \cite{AMPS} was at Page time for an evaporating black hole, while the typical state firewalls of Marolf and Polchinski \cite{Marolf-Polchinski} may arise in large-AdS black holes formed by quick collapse at exponentially late times after complexity has saturated. We would like to emphasize in this section that the immediate regime of applicability of our calculation is in the relatively early (post-scrambling, but pre-Page) times. Therefore we expect our claims in this paper to be relatively uncontroversial. We also expect that clarifying the timescales in the problem will enable us to apply our methods with suitable extra inputs, to later times where the status of firewalls is not yet fully settled. 

From our explicit (analytical and numerical) calculations in the last section, it is clear that the timescale at which the Hartle-Hawking correlator becomes subleading to the fluctuations is about $\sim \mathcal{O}(S_{BH})$. This is natural -- some version of the breakdown of unitarity happens around this timescale for both evaporating and eternal black holes \cite{Maldacena, CK-critical}\footnote{Our goal here is not to be too precise about the $N$ counting. That depends on details. But we would like to distinguish between $\mathcal{O}(\log S_{BH}), \mathcal{O}(S_{BH}^\#)$ and $e^{\mathcal{O}(S_{BH})}$. We are not interested in the precise powers $\#$, or pre-factors at this stage.}. This is the timescale at which we expect non-perturbative effects like new saddles to become significant \cite{Maldacena}. It is worth noting here that non-perturbative $e^{-\mathcal{O}(N)}$ effects may become important at timescales that are $\mathcal{O}(N)$. The latter is what we will call the Page time in our system. Our black holes are implicitly above the Hawking-Page transition, so they don't evaporate away. But this is nonetheless a long enough timescale that one could formulate a version of the information paradox  -- the exact HH correlator undergoes exponential decay and can become arbitrarily small if one waits long enough. This is not what one expects in a unitary theory in finite volume, where the correlator should eventually be characterized by exponentially (in entropy) suppressed non-vanishing fluctuations. We saw that this is indeed what happens in our stretched horizon system.

Note that in our calculation, we started with a typical state. It is the insertion of the first scalar at $t=0$ that allows us to probe the notion of smoothness in a meaningful way using the HH correlator. Our calculation shows that the horizon is smooth once it is perturbed by the first scalar, up to the Page time. Note that if we were to compute 1-point functions in our setting, they will be zero -- not just in the strict large-$N$ limit, but also when there is a finite stretched horizon\footnote{This is true for diagonal expectation values. Off-diagonal correlators can be non-vanishing.}. This is because we are working with a free theory. A free theory is a good model for a generalized free field in the large-$N$ limit, but it may be interesting to include couplings that go as $\frac{1}{\lambda}$ while working in the leading-$N$ approximation, to our set up. These correspond to tree level string corrections. They will allow us to compute non-vanishing 1-point functions, which have recently been used to compute the time to the singularity \cite{Grinberg, David}. 

It has been speculated in \cite{Susskind} that the horizon may be smooth after a perturbation for exponentially long times, ie., up to complexity saturation. Let us discuss some related facts, briefly.

There are two other timescales (other than the Page time) that are of interest when discussing black hole physics. The first is the scrambling time, which scales as $\log S_{BH}$. This is the timescale in which we expect fluctuations to settle down to their thermal values. The Hartle-Hawking correlator is not sensitive to this timescale, because it effectively treats the process of the thermalization after the first scalar field insertion at $t=0$ as instantaneous. So the two point function simply reflects the thermal expectation. In our calculation, where we started with a typical state, this happened because we made the replacement $N \rightarrow \langle N \rangle$ already at $t=0$. Since we have a fully ``mechanical" picture of the dynamics, it should be possible to compute the scrambling time in our set up by not making this approximation and explicitly computing the time to thermalization -- perhaps with some extra inputs, like keeping track of the $J$-dependence of the spectrum. This is clearly of interest, and we hope to come back to it elsewhere. A calculation of this type goes somewhat beyond 't Hooft's original goals because it deals with the {\em process} of thermalization and not just black hole thermodynamics in equilibrium.

Now let us make some comments on timescales that are much larger than $\mathcal{O}(S_{BH})$, all the way up to $e^{\mathcal{O}(S_{BH})}$. Beyond the Page time, the conventional Hartle-Hawking correlator  gets corrected by the requirement of unitarity, and therefore it is not a good probe of the smoothness of the horizon\footnote{\label{probe} It is not entirely clear to us whether the breakdown of the semi-classical Hartle-Hawking correlator at $\mathcal{O}(S_{BH})$ times is to be viewed as a breakdown of the probe we are using or the breakdown of smoothness itself. We will adopt the stand that it is a breakdown of the probe, and that smoothness is intact at $\mathcal{O}(S_{BH})$. This seems to be more in line with the  current conventional wisdom, but it is certainly conceivable that this interpretation is wrong. One reason for thinking that smoothness is preserved is that the growth of the black hole interior stays linear (and consistent with expectations from complexity) even after the Page time.}. A quantity that is sometimes used as an indicator that the horizon is smooth, is the size of Einstein-Rosen bridge in the interior of the eternal black hole. This object grows linearly (indicating the growth of complexity) at least up to exponentially late times, which may be viewed as an indication that the interior is meaningful (and that the horizon is smooth) all the way up to (at least) $e^{\mathcal{O}(S_{BH})}$ times. The latter is the timescale at which complexity saturates. It will be very  interesting to formulate a calculation that is sensitive to these large timescales in our stretched horizon set up. 

Note that our discussion on timescales here has a parallel in the emergence of type III algebras in the large-$N$ limit. A key point is that the half-sided modular translations that were used in \cite{Leutheusser1, Leutheusser2} to probe the interior, exist only in the large-$N$ limit when we can work with generalized free fields. At large timescales like Page time and beyond, one needs a concrete definition of the large-$N$ limit before one can discuss a meaningful notion of the interior. This was discussed in \cite{Vyshnav} where it was  also observed that one can make interior reconstructions state-independent when modular translations are available.  Our results in this paper can be viewed as a more explicit realization of these observations, where we see a smooth horizon again at early times (ie., before the Page time). This is natural from the perspective of type III algebras because part of the motivation for arguing their emergence \cite{Leutheusser2} was the spectral properties of the Hartle-Hawking correlator. Clearly, if Maldacena's information paradox has to be resolved, we need modifications to the HH correlator at Page time. This is precisely what we are finding\footnote{See the very recent paper \cite{Roji1} which also comes to similar conclusions, but phrased as a breakdown of half-sided modular translations at Page time.}. 

We conclude with some brief comments about typical state firewalls. Again, as pointed out in \cite{Vyshnav} it is hard to resolve the question of typical state firewalls within the setting of the emergent type III algebra discussions of \cite{Leutheusser2} because the latter discussions are in the large-$N$ limit. If $N$ is sent to infinity first, then complexity never saturates and we never reach typicality. In the present paper we are working with bulk models for finite-$N$, but the fact that we are ignoring backreaction means that we need new ideas to study late time questions. It will be very interesting to study the trade-off between late times and large-$N$ more carefully. This may help us to identify the precise scaling limit where the smooth horizon is a meaningful idea at finite-$N$ at late times.

\section{Conclusions and Comments}

In this paper, we have partially concretized some ideas that have been around in the zeitgeist, for decades. We have also shown that this naturally fits in with recent observations about emergent algebras in AdS/CFT. The stretched horizon is an old idea \cite{tHooft, STU} but it has always been viewed with a certain air of suspicion\footnote{This is an interesting piece of history, considering the fact that the paper \cite{STU} contains the phrase ``stretched horizon" in the title itself. We believe the discomfort with the stretched horizon arises because one views it as a phenomenological/EFT ingredient, rather than as a piece of the UV complete bulk description (as we have done here). We will clarify what we mean by this, momentarily.}. Our calculation, while not the final word on the matter\footnote{In particular, we have provided strong circumstantial evidence that the stretched horizon is a physical object in the finite-$N$ UV-complete bulk description. Of course, it will be much more satisfying to derive it in a controlled UV complete bulk theory like (a future version of) string theory, rather than postulate it and find evidence for it.}, gives us a plausible way to appreciate the stretched horizon, without feeling too guilty about it. It also provides a relatively precise picture for the phrase ``black hole thermal atmosphere".

Our key point is that if one views the stretched horizon as an ingredient from the finite-$N$ UV-complete description (or more pragmatically, as a UV regulator), there is no immediate contradiction with the principle of equivalence. This is because the principle of equivalence is a feature of low energy bulk EFT\footnote{We suspect that our perspective here may be original, despite the numerous related discussions  in the literature. In \cite{STU} for example, the viewpoint was instead that the stretched horizon is to be viewed as a ``phenomenological description" of the black hole, ``appropriate to a distant observer".}. There is genuine contradiction, only if the {\em effective} correlations of this system are not compatible with those of bulk EFT with a smooth horizon. Our calculation gave strong evidence, that this is {\em not} the case. Despite the presence of the brickwall, we found the Hartle-Hawking correlator of the smooth horizon to emerge, with precisely the correct temperature (and precisely the $e^{-S_{BH}/2}$ suppression of variance). 

\noindent
Let us conclude by making some comments.  
\begin{itemize}
\item Together with the work of \cite{Pradipta}, we have presented evidence that a stretched horizon along with a $J_{cut}$ is enough to fix {\em both} the temperature and entropy, {\em exactly}. In 't Hooft's original calculation, the temperature needed to be specified and the entropy was only fixed up to an unknown number. The precise match was one of the ingredients for us to get the right temperature in the HH correlator, and the correct entropic suppression at finite-$N$.
\item  Complementarity of some form from fuzzballs has been suggested before \cite{Mathur-FuzzComp}. It may be useful to  make fuzzball complementarity \cite{Mathur-NoFire} more precise and try to relate it to our picture here. Note also some recent papers which make comments in the general direction of an emergent smooth horizon from the stretched horizon and normal modes \cite{Nomura, Adepu, Sumit1, Soni}\footnote{See also \cite{Terashima} for earlier comments on brickwalls in AdS/CFT.}. In \cite{Sumit1} a general fluctuation profile on the stretched horizon was assumed, instead of a constant Dirichlet condition. It seems possible that the highly excited typical states on the stretched horizon that we considered in this paper, are another characterization of such a generic profile. It will be good to develop this more concretely. Our discussion is different from these previous discussions, in many ways. Among other things, we are working with a highly excited single-sided pure state on the stretched horizon and not a thermofield double. Our concrete calculation allows us to determine the (emergent) temperature and entropy precisely, and their values are neither put in by hand nor are they fixed only up to numerical pre-factors.
\item  The bulk EFT description differs spectacularly from our model for the UV complete description, starting (about) a Planck length before the horizon. And yet, a remarkable fact about the calculation is that the effective field theory associated to this UV complete description is well-defined, seemingly at and beyond horizon. This is very different from our experience with non-gravitational effective field theories like fluid dynamics. When the UV complete description and the EFT description start departing from each other, usually that indicates that the EFT description has become entirely useless and invalid. This breakdown happens at some UV cutoff where the EFT description is no longer applicable, and we have to resort to a molecular theory (or quantum field theory) rather than fluid dynamics. This is not what seems to be happening at the black hole horizon -- EFT correlators are well-defined and indicate a smooth horizon, and yet the UV complete description (ie., the stretched horizon) differs substantially from it. This raises a possible scenario for fuzzball-like paradigms to work, without doing damage to bulk EFT at the horizon. 
\item It has been known for a while\footnote{This was recently re-iterated in \cite{Zhiboedov}.}, see eg. \cite{Festuccia}, that the Hartle-Hawking correlator contains the complete information about the quasi-normal modes of the black hole. This means that our calculation is implicitly aware of the quasi-normal (ingoing) boundary conditions, even though we started out with a Dirichlet boundary condition. Note that even in the strict $G_N \rightarrow 0$ limit, there is no direct sense in which we can view our boundary condition as ingoing. Yet, the quasi-normal mode information is captured, which is another manifestation of the emergence of smoothness. See \cite{Arnab2} for upcoming discussions about quasi-normal modes from the stretched horizon. 
\item If one views the stretched horizon as a piece of the UV-complete description, it is natural to view complementarity as a feature seen by the low energy probe (which was a scalar field in this paper). This picture suggests that the interior is being ``manufactured" by the probe's effective description. This type of strong complementarity, may perhaps be unsurprising from the point of view that gravity is an effective thermodynamic theory.   
\item Some of our claims in this paper were accomplished using typical states. It will be interesting to go beyond the thermal questions we have considered in this paper, and move more towards chaotic dynamics, scrambling, OTOCs\footnote{Lower point OTOCs  have been used to probe scrambling in both equilibrium and out-of-equilibrium CFTs in the presence of a boundary in \cite{S1, S2, S3}.}, and thermal{\it ization} in general. It may be useful to consider eigenstates themselves and not just the typical states in the microcanonical sliver. Eigenstate thermalization is closely related to chaos -- in our previous papers  \cite{Adepu, Sumit1, Riemann} we noted that the logarithmic dependence of the spectrum on $J$ was crucial for giving signatures of chaos and random matrices. Even though the part of the spectrum responsible for the physics in \cite{Adepu, Sumit1, Riemann} was also the one that controlled the physics in the present paper, we could ignore the $J$ dependence here and still get the thermal physics. It will be very interesting to make progress on more dynamical questions by retaining the $J$-dependence of the spectrum and eigenstates, at least partially.
\item Our calculation naturally suggested a cut-off in the angular quantum number, $J_{cut}$. The way we fixed this quantity was a crucial conceptual way in which our calculation differed from 't Hooft's. It is this difference in our approaches that enabled us to find a precise match of {\em both} the entropy and temperature.  It is very tempting to think that this cut-off in $J$ has stringy origins\footnote{See the papers \cite{Itzhaki, Mertens, Diptarka} which may be relevant here.}. Note in particular that our modes are true normal modes, while 't Hooft's modes were the semi-classically trapped modes behind the angular momentum barrier which were not aware of tunneling. It will be very interesting to understand these questions better, and move the discussion of $J_{cut}$ more towards physics than heuristics.  
\item Finally, let us clarify why we view the results of this paper as evidence for an emergent smooth horizon. What we have shown is that typical excited states on a manifestly ``un-smooth" horizon, lead to the Hartle-Hawking correlator. There are two reasons why this is evidence for emergent smoothness. Firstly, demanding suitably smooth boundary conditions (in terms of Kruskal modes which require the validity of the principle of equivalence at the horizon) is how one usually obtains the Hartle-Hawking correlator. Secondly, the canonical thermal correlator allows a re-writing in terms of the  thermofield double, with which one immediately obtains (see \cite{Leutheusser1, Leutheusser2}) modular translated interior operators, reconstructed interior times, etc. Note that modular translations are available, precisely in the large-$N$ limit. Our results are not yet sufficient to make a clear statement about smoothness of old black holes beyond the Page time -- but they provide evidence for smoothness of not-too-old black holes, in a quantum mechanical setting where there was no smoothness in the microscopic description. 
\end{itemize}

\section{Acknowledgments}  

We thank Pradipta Pathak for numerous discussions, which were crucial for this project. We also thank Souvik Banerjee, Pallab Basu, Diptarka Das, Suman Das, Justin David, Shouvik Datta, Prasad Hegde, Arnab Kundu and Vyshnav Mohan for discussions. The research work of SD is supported by a DST Inspire Faculty Fellowship.

\appendix

\section{Klein-Gordon Inner Product}\label{KGApp}

\noindent For a $D+1$-dimensional spacetime, if we write the metric as 
\begin{align}\label{General metricA}
    ds^{2} = - N(x)^{2}dt^{2} + G_{ij}(x)dx^{i}dx^{j}
\end{align}
then the Klein-Gordon inner product\footnote{See \cite{Crispino} for a clear discussion, whose notations we are following. A slightly dated introduction to QFT in curved space can be found in \cite{CK-Croatia}.} for the modes $f_{A}$ and $f_{B}$ is defined as
\begin{align}\label{KG productA}
\left(f_{A},f_{B}\right)_{KG} = i\int d\vec{x}^{D}\sqrt{G}N^{-1}\left(f_{A}^{*}\partial_{t}f_{B}-f_{B}\partial_{t}f_{A}^{*}\right)
\end{align}
Here $G$ is the determinant of $G_{ij}$. We are working with the 2+1-dimensional BTZ Black hole whose metric is
\begin{align}\label{BTZ metric2A}
ds^2 = -\frac{(r^2-r_{h}^{2})}{L^2}dt^2 + \frac{L^2}{(r^2-r_{h}^{2})}dr^2 + r^2d\psi^2
\end{align}
From the metric (\ref{BTZ metric2A}), we can find out $G$ and $N$.
\begin{align}
    N = \sqrt{\frac{(r^2-r_{h}^{2})}{L^2}} \hspace{0.5cm}; \hspace{0.5cm} G_{ab} =\begin{pmatrix} \frac{L^2}{(r^2-r_{h}^{2})} & 0 \\ 0 & r^2 \end{pmatrix} \hspace{0.5cm}; G \equiv Det(G_{ab}) = \frac{r^{2}L^{2}}{(r^2-r_{h}^2)}
\end{align}
Now using equation (\ref{KG productA}) for the modes given in equation (\ref{Mode}), we can calculate the KG-product:
\begin{align}
&\hspace{-0.8cm}\left(\mathcal{U}_{n,J}(r,t,\psi),\mathcal{U}_{n',J'}(r,t,\psi)\right)_{KG}  = i\int drd\psi \frac{rL^{2}}{(r^2-r_{h}^2)}\Big[ \mathcal{U}_{n,J}^{*}\partial_{t}\mathcal{U}_{n',J'} - \mathcal{U}_{n',J'}\partial_{t}\mathcal{U}_{n,J}^{*}\Big] \\ 
&\hspace{-0.8cm}= i\int drd\psi \frac{L^2}{(r^2-r_{h}^{2})}\Big[ e^{i\omega(n,J)t-i\omega(n',J')t} -e^{-iJ\psi+iJ'\psi}\phi_{n,J}^{*}(r)\phi_{n',J'}(r)(-i\omega(n',J'))- \nonumber \\
&\hspace{5.2cm}-e^{i\omega(n,J)t-i\omega(n',J')t} e^{-iJ\psi+iJ'\psi}\phi_{n,J}^{*}(r)\phi_{n',J'}(r)(i\omega(n,J))\Big] \nonumber\\
&\hspace{-0.8cm}= i\int drd\psi \frac{L^2}{(r^2-r_{h}^{2})}\Big[ e^{i\omega(n,J)t-i\omega(n',J')t} e^{-iJ\psi+iJ'\psi}\phi_{n,J}^{*}(r)\phi_{n',J'}(r)(-i\omega(n',J')-i\omega(n,J)) \Big] \nonumber\\
&\hspace{-0.8cm} = (2\pi) \delta_{J,J'}(\omega(n,J)+\omega(n',J))e^{i(\omega(n,J)-\omega(n',J))t}\Bigg[\int_{r_{0}}^{\infty}dr\frac{L^2}{(r^2-r_{h}^2)}\phi^{*}_{n,J}(r)\phi_{n',J}(r)\Bigg]
\end{align}
In the $4^{th}$ line, we have done the integration over $\psi$. Since modes are normalized using KG product as $(f_A,f_B)_{KG} = \delta_{AB}$, the above equation becomes-
\begin{align}\label{KG solutionA}
\hspace{-4.4mm}\delta_{n,n'}\delta_{J,J'} = (2\pi)\delta_{J,J'}(\omega(n,J)+\omega(n',J))e^{i(\omega(n,J)-\omega(n',J'))t}\Bigg[\int_{r_{0}}^{\infty}\hspace{-1mm}dr\frac{L^2}{(r^2-r_{h}^2)}\phi^{*}_{n,J}(r)\phi_{n',J}(r)\Bigg]
\end{align}
\noindent We demand that 
\bea
\int_{r_{0}}^{\infty}dr\frac{L^2}{(r^2-r_{h}^2)}\phi^{*}_{n,J}(r)\phi_{n',J}(r) = \delta_{n,n'}.
\eea
One thing to note is that the above equation (\ref{KG solutionA}) is only satisfied when our BTZ modes $\mathcal{U}_{n,J}$ contains a factor of $\frac{1}{\sqrt{4\pi\omega_{n,J}}}$. The identity that we have demanded above will be used to find the normalization factor for $\phi_{n,J}(r)$ in Appendix \ref{NormApp}.
\noindent The mode expansion now becomes
\begin{align}
\Phi(r,t,\psi) = \sum_{n,J}\frac{1}{\sqrt{4\pi\omega_{n,J}}}\left(b_{n,J}\mathcal{U}_{n,J}(r,t,\psi)+b^{\dagger}_{n,J}\mathcal{U}^{*}_{n,J}(r,t,\psi)\right)
\end{align}

\section{Stretched Horizon Expansion}\label{StretchExp}

\noindent We do the expansion of the field defined in (\ref{normalized at boundary 2}) around $\epsilon=0$ and get
\begin{align}
\phi_{hor}(r)\approx C_{1}\Bigg[\left(\frac{\epsilon}{r_{h}}\right)^{-\frac{i\omega}{2r_{h}}} P_{1} + \left(\frac{\epsilon}{r_{h}}\right)^{\frac{i\omega}{2r_{h}}}Q_{1}\Bigg]
\end{align}
where\footnote{The reader can re-instate the $L$ dependence in these two expressions via the replacement $\omega \rightarrow \omega L^2$ and $J \rightarrow J L$.}
\begin{align}\label{P1}
P_{1} = \left[\frac{-i2^{-\frac{i\omega}{2r_{h}}}e^{-\frac{\pi}{r_{h}}(\omega+J)}e^{-\frac{i\pi}{2}(1+\nu)}\pi r_{h}^{\frac{1}{2}-\frac{i}{r_{h}}(\omega+J)}\csc h\left(\frac{\pi\omega}{r_{h}}\right)\Gamma\left(\frac{1}{2}(1-\frac{i\omega}{r_{h}}+\frac{iJ}{r_{h}}+\nu)\right)}{\Gamma\left( \frac{iJ}{r_{h}}\right)\Gamma\left(1-\frac{i\omega}{r_{h}}\right)\Gamma\left(\frac{1}{2}\left(1+\frac{i\omega}{r_{h}}-\frac{iJ}{r_{h}}+\nu\right)\right)}  \right]
\end{align}
\begin{align}\label{Q1}
Q_{1} = \left[\frac{i2^{\frac{i\omega}{2r_{h}}}e^{-\frac{\pi}{r_{h}}(\omega+J)}e^{-\frac{i\pi}{2}(1+\nu)}\pi r_{h}^{\frac{1}{2}-\frac{i}{r_{h}}(\omega+J)}\csc h\left(\frac{\pi\omega}{r_{h}}\right)\Gamma\left(\frac{1}{2}(1+\frac{i\omega}{r_{h}}+\frac{iJ}{r_{h}}+\nu)\right)}{\Gamma\left( \frac{iJ}{r_{h}}\right)\Gamma\left(1+\frac{i\omega}{r_{h}}\right)\Gamma\left(\frac{1}{2}\left(1-\frac{i\omega}{r_{h}}-\frac{iJ}{r_{h}}+\nu\right)\right)} \right]
\end{align}
One can directly use (\ref{P1}) and (\ref{Q1}) into (\ref{quantization condition}) to find out the normal modes. But the ratio in (\ref{P1/Q1}) is not directly apparent from (\ref{P1}) and (\ref{Q1}). We convert $P_1$ and $Q_1$ into a more convenient form below. 

In the expression of $P_1$ we multiply and divide by $\Gamma\big(\frac{1}{2}(1+\frac{iL}{r_{h}}(\omega L+J)-\nu)\big)$ and $\Gamma\big(-\frac{iJL}{r_{h}}\big)$. Now using,
\begin{align}
    \Gamma\Big[\frac{1}{2}\Big(1-\frac{iL}{r_{h}}(\omega L-J)+\nu\Big)\Big]\Gamma\Big[\frac{1}{2}\Big(1+\frac{iL}{r_{h}}(\omega L-J)-\nu\Big)\Big] = \frac{2\pi}{e^{-\frac{\pi L}{2r_{h}}(\omega L-J)-\frac{i\pi\nu}{2}}+e^{\frac{\pi L}{2r_{h}}(\omega L-J)+\frac{i\pi\nu}{2}}}
\end{align}
and
\begin{align}
    \Gamma\Big[-\frac{iJL}{r_{h}}\Big]\Gamma\Big[\frac{iJL}{r_{h}}\Big] = \frac{2\pi r_{h}}{JLe^{-\frac{\pi JL}{r_{h}}}\Big(e^{\frac{2\pi JL}{r_{h}}}-1\Big)}
\end{align}
$P_{1}$ reduces to 
\begin{align}\label{final P1}
    P_{1} = -\frac{2^{-\frac{i\omega L^{2}}{2r_{h}}}e^{-\frac{\pi JL}{r_{h}}}(\pi JL)\Big(e^{\frac{2\pi JL}{r_{h}}}-1\Big)r_{h}^{-\frac{1}{2}-\frac{iL}{r_{h}}(\omega L+J)}\text{csch}\Big(\frac{\pi\omega L^{2}}{r{h}}\Big)\Gamma\Big(-\frac{iJL}{r_{h}}\Big)}{\Big(e^{\frac{\pi JL}{r_{h}}}+e^{\pi\big(i\nu+\frac{\omega L^{2}}{r_{h}}\big)}\Big)\Gamma\Big(1-\frac{i\omega L^{2}}{r_{h}}\Big)\Gamma\Big(\frac{1}{2}\Big(1+\frac{iL}{r_{h}}(\omega L-J)+\nu\Big)\Big)\Gamma\Big(\frac{1}{2}\Big(1+\frac{iL}{r_{h}}(\omega L-J)-\nu\Big)\Big)}
\end{align}
For $Q_{1}$, we divide and multiply by $\Gamma\Big(\frac{1}{2}\Big(1-\frac{iL}{r_{h}}(\omega L+J)+\nu\Big)\Big)$ and use 
\begin{align}
    \Gamma\Big[\frac{1}{2}\Big(1+\frac{iL}{r_{h}}(\omega L+J)+\nu\Big)\Big]\Gamma\Big[\frac{1}{2}\Big(1-\frac{iL}{r_{h}}(\omega L+J)-\nu\Big)\Big]= \frac{2\pi}{e^{-\frac{\pi L}{2r_{h}}(\omega L+J)}e^{-\frac{i\pi\nu}{2}}\Big(e^{i\pi\nu}+e^{\frac{\pi L}{r_{h}}(\omega L+J)}\Big)}
\end{align}
Then $Q_{1}$ reduces to 
\begin{align}\label{final Q1}
    Q_{1} = \frac{(-1)^{\frac{i\omega L^{2}}{r_{h}}}2^{\frac{i\omega L^{2}}{2r_{h}}+1}e^{\frac{2\pi\omega L^{2}}{r_{h}}}\pi^{2}r_{h}^{\frac{1}{2}-\frac{i\omega L^{2}}{r_{h}}-\frac{iJL}{r_{h}}}\Big(coth\Big(\frac{\pi\omega L^{2}}{r_{h}}\Big)-1\Big)} {\Big(e^{i\pi\nu}+e^{\frac{\pi L}{r_{h}}(\omega L+J)}\Big)\Gamma\Big(\frac{iJL}{r{h}}\Big)\Gamma\Big(1+\frac{i\omega L^{2}}{r_{h}}\Big)\Gamma\Big(\frac{1}{2}\Big(1-\frac{iL}{r_{h}}(\omega L+J)+\nu\Big)\Big)\Gamma\Big(\frac{1}{2}\Big(1-\frac{iL}{r_{h}}(\omega L+J)-\nu\Big)\Big)}
\end{align}
We use (\ref{final P1}) and (\ref{final Q1}) forms of the $P_{1}$ and $Q_{1}$ to write the ratio $\frac{P_{1}}{Q_{1}}$ in (\ref{P1/Q1}).

\section{Normalization of the Field $\phi_{n,J}(r)$}\label{NormApp}

As we noted in appendix \ref{KGApp} the Klein-Gordon inner product for BTZ normal modes leads to the following relation
\begin{align}\label{KGB}
\int^{\infty}_{r_{h}+\epsilon}dr \frac{L^2}{r^{2}-r_{h}^{2}}\phi_{n,J}\phi^{*}_{n',J} = \delta_{n,n'}
\end{align}
Here $\epsilon$ is the location of the stretched horizon which is of the order Planck length ($\epsilon \ll r_{h}$). The normalizability of the modes near boundary leads us to the following expression\footnote{Note that here we denote $\omega L^{2}/r_{h} \rightarrow \omega$ and $J L/r_{h} \rightarrow J$. Also, we are writing $\omega_{n,J}$ to be $\omega$ for convenience -- i.e., the $\omega$ should be understood as quantized modes.}
\begin{align}\label{phiB}
&\phi_{n,J}(r) = C_{n,J} S r^{\frac{1}{2}-i\frac{JL}{r_{h}}} (r^{2}-r_{h}^{2})^{-\frac{i\omega L^{2}}{2r_{h}}}e^{-\frac{i\pi}{2}(1-i\frac{\omega L^{2}}{r_{h}}-i\frac{JL}{r_{h}}+\nu)} \left(\frac{r}{r_{h}}\right)^{-(1-i\frac{\omega L^{2}}{r_{h}}-i\frac{JL}{r_{h}}+\nu)} 
 {}_2F_{1}\left(\alpha,\beta;\Delta;y\right)
\end{align}
Here $C_{n,J}$ is some undetermined constant which we would like to fix using (\ref{KGB}). However, from (\ref{KGB}) one can only fix $|C_{n,J}|$. To fix the ${\rm Arg}(C_{n,J})$, we will use the reality condition of the modes, i.e.
\begin{align}\label{realityB}
\phi_{n,J}(r)=\phi^{*}_{n,J}(r)
\end{align}
Putting (\ref{phiB}) in (\ref{KGB}) we get\footnote{We will only be interested in extracting the normalization from \eqref{KGB} in what follows. But let us emphasize that the ortho$gonality$ property of \eqref{KGB} also holds at finite $\epsilon$ (and $not$ merely in a suitable $\epsilon \rightarrow 0$ limit). This is guaranteed by the fact that the radial equation here is a conventional Sturm-Liouville problem \cite{Sturm}. We believe the resulting hypergeometric integral identities are new (at least, we have not been able to find them in the standard sources). The present Appendix can be viewed as an approximate determination of the normalization of that integral. For spherical hole boundary conditions in flat 2+1 d space, we have checked that the analogous orthonormality conditions lead to Bessel function integrals which are known \cite{BesselSturm}. 
},
\begin{align}\label{int1B}
&|C_{n,J}|^{2}e^{-\frac{\pi L}{r_{h}}(\omega L+J)}r_{h}^{2(1+\nu+m+n)}     SS^{*}\sum_{m}\frac{\left(\alpha\right)_{m}\left(\beta\right)_{m}}{(\Delta)_{m}m!}\sum_{n}\frac{\left(\alpha^{*}\right)_{n}
\left(\beta^{*}\right)_{n}}{(\Delta)_{n}n!}
\int^{\infty}_{r_{h}+\epsilon} dr \frac{r^{-1-2(\nu+m+n)}}{r^{2}-r_{h}^{2}} = \frac{1}{L^2}
\end{align}
Here we have used\footnote{Most of our hypergeometric and related identities are taken from Gradshteyn and Ryzhik \cite{Gradshteyn}.} the expansion of hypergeometric function for $|z|<1$\footnote{Note that here $z=\frac{r_{h}}{r}$ and $r>r_{h}$ naturally induces such expansion at $|z|<1$. If our integration range includes $r_{h}$, we could not perform this expansion.}
\begin{align}
{}_2F_{1}(a,b,c,z) = \sum_{m} \frac{(a)_{m}(b)_{m}}{(c)_{m}m!}z^{m}
\end{align}
The integration of the radial part of the above expression yields
\begin{align}\label{normalization integration}
&\lim_{\Lambda \rightarrow \infty} \int^{\Lambda}_{r_{h}+\epsilon} dr \frac{r^{-1-2(m+n+\nu)}}{r^{2}-r_{h}^{2}} = \lim_{\Lambda\rightarrow\infty} \frac{1}{2(1+\nu+m+n)} \left((r_{h}+\epsilon)\Lambda\right)^{-2(1+\nu+m+n)}\times\\
&\hspace{6cm}\times\Big[ F_{1}\Lambda^{2(1+\nu+m+n)}-F_{2}(r_{h}+\epsilon)^{2(1+\nu+m+n)}\Big] \nonumber \\
&\hspace{4.6cm} = \frac{1}{2(1+\nu+m+n)} (r_{h}+\epsilon)^{-2(1+\nu+m+n)} F_{1}
\end{align}
Where, 
\begin{align}
    F_{1} = {}_2F_{1}\left(1,1+m+n+\nu,2+m+n+\nu,\frac{r_{h}^{2}}{(r_{h}+\epsilon)^{2}}\right)
\end{align}
and 
\begin{align}
    F_{2} = {}_2F_{1}\left(1,1+m+n+\nu,2+m+n+\nu,\frac{r_{h}^{2}}{\Lambda^{2}}\right)
\end{align}
In the limit $\Lambda\rightarrow\infty$, the second term inside the square bracket in (\ref{normalization integration}) vanishes. The first hypergeometric function inside the square bracket can be written as
\bea
F_{1}& \approx &{}_2F_1(1,1+m+n+\nu,2+m+n+\nu,1-\frac{2\epsilon}{r_{h}})\\
&\approx & -(1+\nu+m+n)\log\Big(\frac{2\epsilon}{r_{h}}\Big)
\eea
Thus we have,
\begin{align}
\lim_{\Lambda \rightarrow \infty} \int^{\Lambda}_{r_{h}+\epsilon} dr \frac{r^{-1-2(m+n+\nu)}}{r^{2}-r_{h}^{2}} = -\frac{1}{2}\log(\frac{2\epsilon}{r_{h}}) (r_{h}+\epsilon)^{-2(1+\nu+m+n)}. 
\end{align}
Plugging this back into (\ref{int1B}) and expanding around small $\epsilon$ we have,
\begin{align}
L^2 M|C_{n,J}|^{2}\sum_{m}\frac{\left(\alpha\right)_{m}\left(\beta\right)_{m}}{(\Delta)_{m}m!}\left(1-\frac{2\epsilon}{r_{h}}\right)^{m}\sum_{n}\frac{\left(\alpha^{*}\right)_{n}\left(\beta^{*}\right)_{n}}{(\Delta)_{n}n!}\left(1-\frac{2\epsilon}{r_{h}}\right)^{n}= 1
\end{align}
To avoid the ugliness of the expression, we define $M$ as
\begin{align}
M=- \frac{SS^{*}}{2}\log\Big(\frac{2\epsilon}{r_{h}}\Big)e^{-\frac{\pi L}{r_{h}}(\omega L+J)} \left(1-\frac{\epsilon}{r_{h}}\right)^{2+2\nu}
\end{align}
Again using $\sum_{m} \frac{(a)_{m}(b)_{m}}{(c)_{m}m!}(1-\epsilon)^{m}= {}_{2}F_{1}(a,b,c,1-\epsilon)$, we obtain
\begin{align}\label{CB}
& L^2 M|C_{n,J}|^{2}{}_2F_{1}\left(\alpha,\beta;\Delta;1-\frac{2\epsilon}{r_{h}}\right)
{}_2F_{1}\left(\alpha^{*},\beta^{*};\Delta;1-\frac{2\epsilon}{r_{h}}\right) = 1 .
\end{align}
Now we will use the following hypergeometric identity which applies when $Re(c)=Re(a+b)$ and $c \neq (a+b)$ and $\epsilon \ll 1$,
\begin{align}\label{Hypergeometric identityB}
{}_{2}F_{1}(a,b,c,1-\epsilon) = \epsilon^{c-a-b}\frac{\Gamma(c)\Gamma(a+b-c)}{\Gamma(a)\Gamma(b)}+ \frac{\Gamma(c)\Gamma(c-a-b)}{\Gamma(c-a)\Gamma(c-b)}.
\end{align}
Using (\ref{Hypergeometric identityB}), the product of two hypergeometric functions in (\ref{CB}) becomes
\begin{align}\label{hyper prodB}
&\nonumber {}_2F_{1}\left(\alpha,\beta;\Delta;1-\frac{2\epsilon}{r_{h}}\right)
{}_2F_{1}\left(\alpha^{*},\beta^{*};\Delta;1-\frac{2\epsilon}{r_{h}}\right) = 2\Gamma^2(\Delta)T^{-2} +\frac{\Gamma^{2}(\Delta)\Gamma^2(\frac{i\omega L^{2}}{r_{h}})\Big(\frac{2\epsilon}{r_{h}}\Big)^{-\frac{i\omega L^{2}}{r_{h}}}}{\Gamma^{2}(\alpha^{*})\Gamma^{2}(\beta^{*})}
+ \nonumber \\
&\hspace{10cm}+\frac{\Gamma^{2}(\Delta)\Gamma^2(-\frac{i\omega L^{2}}{r_{h}})\Big(\frac{2\epsilon}{r_{h}}\Big)^{\frac{i\omega L^{2}}{r_{h}}}}{\Gamma^{2}(\alpha)\Gamma^{2}(\beta)}
\end{align}
This simplifies to
\begin{align}
{}_2F_{1}\left(\alpha,\beta;\Delta;1-\frac{2\epsilon}{r_{h}}\right)
{}_2F_{1}\left(\alpha^{*},\beta^{*};\Delta;1-\frac{2\epsilon}{r_{h}}\right) = \frac{2\Gamma^{2}(\Delta)\Gamma(\frac{i\omega L^{2}}{r_{h}})\Gamma(-\frac{i\omega L^{2}}{r_{h}})(1+A)}{\Gamma\left(\beta^{*}\right)\Gamma\left(\alpha\right)\Gamma\left(\alpha^{*}\right)\Gamma\left(\beta\right)}
\end{align}
where
\begin{align}\label{AB}
&A \equiv \frac{1}{2}\left(\frac{2\epsilon}{r_{h}}\right)^{-\frac{i\omega L^{2}}{r_{h}}}\frac{\Gamma(-\frac{i\omega L^{2}}{r_{h}}) \Gamma \left(\beta^{*} \right) \Gamma\left(\alpha^{*}\right)}{\Gamma(\frac{i\omega L^{2}}{r_{h}})\Gamma\left (\beta\right)\Gamma\left(\alpha\right)}\left[ \frac{\Gamma^{2}(\frac{i\omega L^{2}}{r_{h}})\Gamma^{2}\left(\alpha\right)\Gamma^{2}\left(\beta\right)}{\Gamma^{2}(-\frac{i\omega L^{2}}{r_{h}})\Gamma^{2}\left(\alpha^{*}\right)\Gamma^{2}
\left(\beta^{*}\right)} + \left(\frac{2\epsilon}{r_{h}}\right)^{\frac{2i\omega L^{2}}{r_{h}}}\right]
\end{align}
\noindent Finally by putting (\ref{hyper prodB}) in (\ref{CB}) we get
\begin{align}\label{final modB}
&L\ |C_{n,J}| = \left(-(1+A)\log\left(\frac{2\epsilon}{r_{h}}\right)\right)^{-\frac{1}{2}} \left(1-\frac{\epsilon}{r_{h}}\right)^{-1-\nu}\left(\frac{\Gamma(\frac{iJL}{r_{h}})\Gamma(-\frac{iJL}{r_{h}})}{\Gamma(\frac{i\omega L^{2}}{r_{h}})\Gamma(-\frac{i\omega L^{2}}{r_{h}})}\right)^{\frac{1}{2}}e^{\frac{\pi L}{2}(\omega L+J)}
\end{align}
\noindent Now our task is to find the ${\rm Arg}(C_{n,J})$ from the reality condition (\ref{realityB}). Denoting $C_{n,J}=|C_{n,J}|e^{i\theta}$, (\ref{realityB}) yields
\begin{align}
\hspace{-4mm}e^{-2i\theta} = r^{-\frac{2iJ}{r_{h}}}(r^{2}-r_{h}^{2})^{-\frac{i\omega L^{2}}{r_{h}}}e^{-i\pi\Delta}\left(\frac{r}{r_{h}}\right)^{\hspace{-1mm}\frac{2iL}{r_{h}}(\omega L+J)}   \frac{\Gamma\left(\beta^{*}\right)\Gamma\left(\alpha\right)\Gamma(-iJL/r_h)}{\Gamma\left (\beta\right) \Gamma\left(\alpha^{*} \right)\Gamma(iJL/r_h)}\frac{{}_2F_{1}\left(\alpha,\beta;\Delta;y\right)}{{}_2F_{1}\left(\alpha^{*},\beta^{*};\Delta;y\right)}
\end{align}
To find the ratio of the two hypergeometric functions given in the above equation, we use the following hypergeometric identity
\begin{align}
{}_{2}F_{1}(a,b,c,z)=(1-z)^{c-a-b}{}_{2}F_{1}(c-a,c-b,c,z),
\end{align}
which gives,
\begin{align}
\frac{{}_2F_{1}\left(\alpha,\beta;\Delta;y\right)}{{}_2F_{1}\left(\alpha^{*},\beta^{*};\Delta;y\right)} = \left(1-\frac{r_{h}^{2}}{r^{2}}\right)^{\frac{i\omega L^{2}}{r_{h}}}
\end{align}
The expression for $e^{i\theta}$ thus reduces to
\begin{align}\label{final argB}
\hspace{-4mm}e^{i\theta} &= r^{\frac{iJL}{r_{h}}}(r^{2}-r_{h}^{2})^{\frac{i\omega L^{2}}{2r_{h}}}e^{\frac{i\pi}{2}\Delta}\left(\frac{r}{r_{h}}\right)^{-\frac{iL}{r_{h}}(\omega L+J)} \left(1-\frac{r_{h}^{2}}{r^{2}}\right)^{-\frac{i\omega L^{2}}{2r{h}}}\left(\frac{\Gamma\left(\beta
\right)\Gamma\left(\alpha^{*}\right)\Gamma(\frac{iJL}{r_{h}})}{\Gamma\left(
\beta^{*}\right)\Gamma\left(\alpha\right)\Gamma(-\frac{iJL}{r_{h}})}
\right)^{\frac{1}{2}} 
\\  &= e^{\frac{i\pi}{2}\Delta} \; r_{h}^{\frac{iL}{r_{h}}(\omega L+J)} \left(\frac{\Gamma(\beta)\Gamma(\alpha^{*})\Gamma(\frac{iJL}{r_{h}})}{\Gamma(\beta^{*})\Gamma(\alpha)\Gamma(-\frac{iJL}{r_{h}})}\right)^{\frac{1}{2}}
\end{align}
Note that this final expression is $r$-independent.
The final form of the normalization is 
\begin{align}
    C_{n,J}\equiv|C_{n,J}|e^{i\theta}
\end{align}
which can be written as
\begin{align}\label{final normalizationB}
L\ C_{n,J}=\sqrt{K}r_{h}^{\frac{iL}{r_{h}}(\omega L+J)}e^{\frac{i\pi}{2}(1-\frac{i\omega L^{2}}{r_{h}}-\frac{iJL}{r_{h}}+\nu)}\left[\frac{\Gamma\left(\beta\right)\Gamma
\left(\alpha^{*}\right)\Gamma\left(\frac{iJL}{r_{h}}\right)^{2}}{\Gamma\left
(\beta^{*}\right)\Gamma\left(\alpha\right)\Gamma(\frac{i\omega L^{2}}{r_{h}})
\Gamma(-\frac{i\omega L^{2}}{r_{h}})}\right]^{\frac{1}{2}}
\end{align}
Previously, we defined $K$ to be $\frac{1}{\log\Big(\frac{r_{h}}{2\epsilon}\Big)}\Big(1-\frac{\epsilon}{r_{h}}\Big)^{-2\Delta}$.
Finally using (\ref{final modB}) and (\ref{final argB}), the final expression of (\ref{phiB}) reduces to
\begin{align}\label{phi finalB}
L \ \phi_{n,J}(r) = \frac{T\sqrt{K}}{\sqrt{(1+A)}}r^{-\frac{1}{2}-\nu}\frac{r_{h}^{1+\nu}}{\Gamma(\Delta)}\left(1-\frac{r_{h}^{2}}{r^{2}}\right)^{-\frac{i\omega L^{2}}{2r_{h}}} {}_{2}F_{1}\left(\alpha,\beta;\Delta;y\right)
\end{align}

Note that the expression (\ref{final modB}) of $|C_{n,J}|$ has a phase factor due to the non-vanishing of $A$, so it is not quite the correct answer. But it turns out that it can be made acceptable if we demand certain relations between $\omega$ and $J$. For $\nu=1$, let us write
\begin{align}
&A = \frac{1}{2}\left(\frac{2\epsilon}{r_{h}}\right)^{-\frac{i\omega L^{2}}{r_{h}}}\frac{\Gamma(-\frac{i\omega L^{2}}{r_{h}}) \Gamma \left(\beta^{*} \right) \Gamma\left(\alpha^{*}\right)}{\Gamma(\frac{i\omega L^{2}}{r_{h}})\Gamma\left (\beta\right)\Gamma\left(\alpha\right)}\left[ \frac{\Gamma^{2}(\frac{i\omega L^{2}}{r_{h}})\Gamma^{2}\left(\alpha\right)\Gamma^{2}\left(\beta\right)}{\Gamma^{2}(-\frac{i\omega L^{2}}{r_{h}})\Gamma^2\left(\alpha^{*}\right)\Gamma^2
\left(\beta^{*}\right)} + \left(\frac{2\epsilon}{r_{h}}\right)^{\frac{2i\omega L^{2}}{r_{h}}}\right] \nonumber \\
&= \frac{1}{2}\left(\frac{2\epsilon}{r_{h}}\right)^{-\frac{i\omega L^{2}}{r_{h}}}\frac{\Gamma(-\frac{i\omega L^{2}}{r_{h}}) \Gamma \left(\beta^{*} \right) \Gamma\left(\alpha^{*}\right)}{\Gamma(\frac{i\omega L^{2}}{r_{h}})\Gamma\left (\beta\right)\Gamma\left(\alpha\right)} \times \nonumber \\
&\times\left[\exp\left(4iArg(\Gamma(\frac{i\omega L^{2}}{r_{h}}))+4iArg(\Gamma(-\frac{iL}{2r_{h}}(\omega L+J)))+ 4iArg(\Gamma(\frac{iL}{2r_{h}}(J-\omega L)))\right)
+ \Big(\frac{2\epsilon}{r_{h}}\Big)^{\frac{2i\omega L^{2}}{r_{h}}}\right]
\end{align}
In other words, $A$ automatically vanishes by (\ref{dispersion eqtn}). Remarkably, the Dirichlet boundary condition at the stretched horizon removes the fictitious phase factor in $|C_{n,J}|$. This is also true for massive field with $\nu \neq 1$ which one might check. 

One can see that the normalized expression of the scalar field which we obtain here, is similar (up to some $\epsilon$ dependent terms) to the usual BTZ solution determined using free wave boundary condition in tortoise coordinates \cite{Festuccia, Yang}. The latter is the conventional smooth horizon boundary condition for the eternal black hole, at future and past horizons. One thing to note here is that the normalization not only differs by some $\epsilon$-dependent factors but also by some power of $r_{h}$. This change in the dependence of $r_{h}$ in the Hartle-Hawking case is also nice because it leads to the $r_{h}$-independent boundary correlator that exactly matches with the boundary limit of the stretched horizon correlator in the $\epsilon \rightarrow 0$ limit. This structural similarity between the two solutions arises because close to the stretched horizon, the normalizable modes can be approximated by free waves in tortoise coordinates -- there is an emergent Rindler universality near the horizon. 

The claim then is that fixing $\phi$ using free wave boundary conditions in tortoise coordinates at the horizon (without using K-G norm) leads to a structure very close to fixing $\phi$ using the orthonormality of K-G inner product with a Dirichlet boundary condition at the stretched horizon. The difference lies in some overall factors like $\log(\frac{2\epsilon}{r_{h}})$ and $\left(1-\frac{\epsilon}{r_{h}}\right)^{-1-\nu}$), as well as in the fact that the $\omega$'s are quantized when there is a stretched horizon. Both features will play crucial roles when we compute the effective correlators in the $\epsilon \rightarrow 0$ limit, in connecting with the Hartle-Hawking result.

In any event, the normalized real radial modes are finally:
\begin{align}\label{phi final2B}
&\phi_{n,J}(r) = \frac{T}{L}\sqrt{K}r^{-\frac{1}{2}-\nu}\frac{r_{h}^{1+\nu}}{\Gamma(\Delta)}\left(1-\frac{r_{h}^{2}}{r^{2}}\right)^{-\frac{i\omega L^{2}}{2r_{h}}} {}_{2}F_{1}\left(\alpha,\beta;\Delta;y\right)
\end{align}

\end{document}